\documentclass[sigconf,nonacm]{aamas}
\usepackage{balance}
\makeatletter
\@ACM@balancefalse
\makeatother

\setcopyright{none}
\acmSubmissionID{}
\submissionType{}
\acmConference[]{}{}{}{}

\usepackage{amsmath}
\usepackage{pifont}
\usepackage{array}
\usepackage{multirow}
\usepackage{float}
\usepackage{placeins}
\usepackage{flafter}
\usepackage{stfloats}
\usepackage{listings}

\newcommand{\bgebase}{\texttt{BAAI/\allowbreak bge-base-\allowbreak en-v1.5}}
\newcolumntype{L}[1]{>{\raggedright\arraybackslash}p{#1}}
\makeatletter
\let\aamasoriginaltitlefont\@titlefont
\renewcommand{\@titlefont}{\aamasoriginaltitlefont\fontsize{14.5}{18}\selectfont}
\makeatother
\title[Is Serving-Time Language-Model Inference Necessary for Recommendation?]{Is Serving-Time Language-Model Inference Necessary for Recommendation?\texorpdfstring{\\}{ }
Evidence from Generative and Agentic Settings}
\author{Xiang Chen}
\authornote{These authors contributed equally to this work.}
\affiliation{\institution{Tongji University}\country{China}}
\author{Zhongke He}
\authornotemark[1]
\affiliation{\institution{Dalian University of Technology}\country{China}}
\author{Hongwei Zeng}
\authornotemark[1]
\affiliation{\institution{University of Chinese Academy of Sciences}\country{China}}
\author{Kai Deng}

\affiliation{\institution{Beihang University}\country{China}}
\author{Long Wang}

\affiliation{\institution{Xinjiang College of Science \& Technology}\country{China}}
\author{Kehua Yang}

\affiliation{\institution{Hunan University}\country{China}}
\author{Kong Wang}
\authornote{Corresponding author.}
\affiliation{\institution{Hunan University}\country{China}}
\renewcommand{\shortauthors}{Xiang Chen et al.}
\makeatletter
\renewcommand{\@mkauthors}{%
  \global\setbox\mktitle@bx=\vbox{%
    \hsize=\textwidth
    \unvbox\mktitle@bx\par\medskip
    \centering
    {\large
      Xiang Chen\textsuperscript{3,*}\quad
      Zhongke He\textsuperscript{2,*}\quad
      Hongwei Zeng\textsuperscript{4,*}\\[4pt]
      Kai Deng\textsuperscript{5}\quad
      Long Wang\textsuperscript{6}\quad
      Kehua Yang\textsuperscript{1}\quad
      Kong Wang\textsuperscript{1,\textdagger}\par}
    \smallskip
    {\normalsize
      \textsuperscript{1}Hunan University\qquad
      \textsuperscript{2}Dalian University of Technology\\[2pt]
      \textsuperscript{3}Tongji University\qquad
      \textsuperscript{4}University of Chinese Academy of Sciences\\[2pt]
      \textsuperscript{5}Beihang University\qquad
      \textsuperscript{6}Xinjiang College of Science \& Technology\par}
    \medskip
  }%
}
\makeatother


\begin{abstract}
Language models (LMs) are increasingly used in recommender systems to generate recommendations or coordinate recommendation tools. However, does their strong performance imply that serving-time LM inference is necessary? We challenge this seemingly reasonable but unverified assumption. To investigate this question, we introduce \textbf{recovery-based mechanism auditing}, which tests whether independent alternatives can achieve the accuracy of LM-based recommender systems without the audited inference mechanisms under controlled information and evaluation conditions. For generative recommendation, we propose \textbf{LIME-Rec}, which combines sequential, collaborative, and offline semantic information without autoregressive LM decoding. On three Amazon Reviews datasets---Beauty, Toys, and Sports---LIME-Rec achieves the highest reported mean in all 12 dataset--metric combinations, with up to a \textbf{15.2\% relative improvement in Recall@10} over GRAM, an LM-based generative recommender. For agentic recommendation, we propose \textbf{DART-Rec}, which replaces LM-based tool control with deterministic fusion and fixed expert selection. On 1,000 Beauty users, given the same candidate items and expert scores, deterministic fusion achieves a \textbf{16.7\% relative improvement in Recall@10} over a fine-tuned LM controller. With at most two tool calls, fixed expert selection matches the LM controllers' Hit@10 outcomes for every user in \textbf{17 of 18 comparisons}. These findings demonstrate that strong recommendation performance does not necessarily require serving-time LM inference: under the evaluated task and information conditions, alternatives without the specified inference mechanisms can match or exceed the accuracy of their LM-based counterparts. To support reproducibility, code is available at \url{https://github.com/Double-wk/LIME-Rec}.
\end{abstract}

\begin{document}
\raggedbottom
\settopmatter{printccs=true,printacmref=false}
\maketitle
\pagestyle{plain}
\thispagestyle{plain}

\section{Introduction}

Language models (LMs) increasingly generate recommendations and coordinate recommendation tools~\cite{Geng2022P5,Bao2023TallRec,Liao2024LLaRA,yao2023react,schick2023toolformer}. Their strong results invite a natural assumption: serving-time LM inference is essential to high recommendation accuracy. Yet system-level performance alone cannot establish that a particular inference mechanism is necessary~\cite{Dacrema2019Progress,rudin2024rashomon}. The same recommendation signals may support an alternative system without serving-time LM inference. We therefore ask: \textbf{Can LM-based recommendation accuracy be attained without serving-time LM inference?}

The distinction matters because a recommendation pipeline can use language models at different stages. Item semantics may be encoded once offline, while the online system ranks a fixed catalog from a user's interaction history. In that setting, a successful LM-based system may benefit from informative representations or strong candidate retrieval even when its serving-time generation or control procedure is replaceable. The practical question is therefore which information must be retained and which online operations are needed to attain the observed ranking accuracy. This question is narrower than whether language models are useful for recommendation as a whole.

Baseline comparisons typically report the accuracy of complete systems~\cite{Dacrema2019Progress}, while ablations measure dependence on components within one system~\cite{Meyes2019Ablation}. What remains unclear is whether an independent alternative can reach the target accuracy while excluding serving-time LM inference. We introduce \textbf{recovery-based mechanism auditing} to test this directly. An audit specifies which mechanism the alternative excludes, what information remains available, and how performance is evaluated. Recovery occurs when the alternative reaches the target accuracy within a stated loss tolerance.

In \textbf{generative recommendation}, we audit autoregressive LM decoding of item identifiers. Our alternative, \textbf{LIME-Rec}, combines sequential, collaborative, and offline semantic signals to rank items without serving-time LM decoding. On Amazon Reviews 2014~\cite{McAuley2015Amazon} Beauty, Toys, and Sports, we compare it with GRAM~\cite{gram}, TIGER~\cite{Rajput2023TIGER}, and LIGER~\cite{liger}. LIME-Rec shows statistically supported superiority in eight of nine target--dataset comparisons and up to a \textbf{15.2\% relative Recall@10 (R@10) gain} over GRAM. The remaining GRAM--Sports comparison supports a weaker plug-in sensitivity result (Section~\ref{sec:recovery-statistics}).

\begin{figure*}[t]
    \centering
    \includegraphics[width=\textwidth]{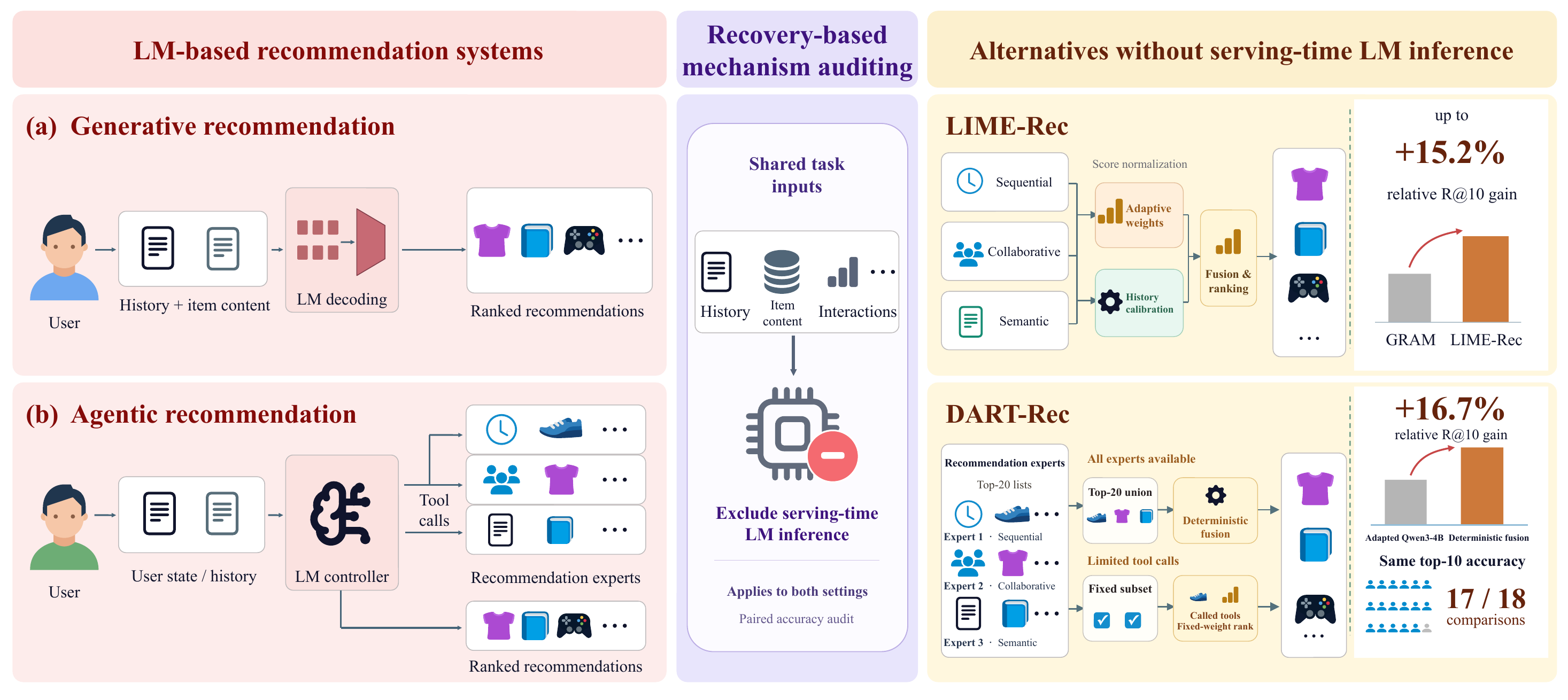}
    \caption{Auditing LM inference in generation and tool control. LIME-Rec ranks items without autoregressive LM decoding; DART-Rec uses deterministic fusion or fixed expert selection. On Toys, LIME-Rec gains 15.2\% in Recall@10 over GRAM. On Beauty, deterministic fusion gains 16.7\% over an adapted LM controller in shared-pool ranking; fixed selection matches per-user Hit@10 in 17 of 18 limited-call comparisons.}
    \Description{
    Overview of two recovery-based mechanism audits. In generative
    recommendation, serving-time LM decoding is replaced by LIME-Rec,
    which combines sequential, collaborative, and semantic evidence and
    achieves up to a 15.2 percent relative R@10 gain over GRAM. In
    agentic recommendation, DART-Rec replaces LM-based tool control with
    deterministic fusion when all expert outputs are available and fixed
    expert selection when tool calls are limited. Deterministic fusion
    reaches R@10 0.112 versus 0.096 for an adapted Qwen3-4B controller
    on the 1,000-user Beauty evaluation with all expert outputs available,
    a 16.7 percent relative gain, while fixed expert
    selection produces
    identical user-level Hit@10 outcomes in 17
    of 18 dataset--seed--controller comparisons.
    }
    \label{fig:overview}
\end{figure*}

In \textbf{agentic recommendation}, we audit LM-based control of recommendation tools. \textbf{DART-Rec} replaces LM control with deterministic fusion when all expert outputs are available and fixed expert selection when tool calls are limited. The audit tests tool selection and ranking in a controlled setting (Figure~\ref{fig:overview}).

On a fixed sample of 1,000 Beauty users, deterministic fusion improves shared-pool Recall@10 by \textbf{16.7\%} over an LM controller adapted to imitate fusion rankings. With limited tool access, fixed selection produces identical per-user binary Hit@10 outcomes in \textbf{17 of 18 comparisons} with LM controllers. Alongside the generative results, these findings identify settings where the audited LM mechanisms are not required to attain the observed accuracy. We also examine where recovery is weaker, including Yelp and longer-history Sports users.

Our contributions are threefold.

\textbf{(1)} We introduce recovery-based mechanism auditing to test whether target accuracy is attainable without a specified serving-time LM mechanism.

\textbf{(2)} We develop LIME-Rec and DART-Rec to test LM decoding and LM-based tool control, respectively.

\textbf{(3)} We evaluate recovery in generative and agentic recommendation and examine where it weakens across datasets and user histories.

\section{Related Work}
\subsection{LM Generation and Tool Control}
In generative recommendation, LMs can autoregressively decode item identifiers. Semantic-ID methods represent item semantics with discrete identifiers, while later approaches incorporate additional semantic, lexical, or retrieval signals~\cite{gram,Rajput2023TIGER,liger}. These methods differ in design but retain LM decoding at serving time.

In agentic recommendation, LMs select tools or coordinate their outputs, drawing on frameworks that combine reasoning, actions, and observations~\cite{huang2024recagent,yao2023react}. In both settings, accuracy depends on more than LM inference: item representations and retrieval support generation, while expert outputs support tool control. End-to-end performance therefore cannot establish whether serving-time LM inference is necessary.

\subsection{Attribution and Mechanism Necessity}
Prior studies show that recommendation gains may diminish under stronger baselines~\cite{Dacrema2019Progress}, while agent evaluations can be affected by system design and cost~\cite{Kapoor2024Agents}. These findings highlight the importance of distinguishing system performance from the contribution of individual mechanisms.

Existing methods address different aspects of this problem. Ablation measures dependence on components within a system~\cite{Meyes2019Ablation}. Rashomon analysis examines alternative models with comparable performance~\cite{rudin2024rashomon}. Non-inferiority testing evaluates performance against a predefined loss margin~\cite{wellek2010equivalence}. However, none of these methods alone requires an independent alternative to exclude a specified inference mechanism.

Studies of fixed agent workflows may also retain LM inference~\cite{Xia2024Agentless}. In contrast, recovery-based mechanism auditing explicitly excludes the audited serving-time mechanism and tests whether an independent alternative can attain target accuracy under declared information and evaluation conditions. Successful recovery demonstrates that the audited mechanism is not required to attain target accuracy under those conditions.

\section{Recovery-Based Mechanism Auditing}
\label{sec:audit-framework}

We test \emph{attainability}: whether an independent alternative can attain an LM-based recommender $A$'s accuracy without its serving-time mechanism $m$. Recovery establishes conditional existence of such an alternative, not the causal effect of deleting $m$ from $A$.

An audit must therefore specify three ingredients: the mechanism to be excluded, the information available to the alternative, and a common evaluation protocol. Let $B=(B_{\mathrm{task}},B_{\mathrm{repr}})$ describe the information conditions and let $\Pi$ denote the evaluation protocol. We write $\mathcal C_{\neg m}(B,\Pi)$ for alternatives that exclude $m$ while satisfying the declared conditions. For a dataset $D$, let $U(A;D)$ be the target's performance on the primary metric. An alternative $R\in\mathcal C_{\neg m}(B,\Pi)$ \emph{recovers} target performance within an absolute tolerance $\delta\geq0$ if
\begin{equation}
U(R;D)\geq U(A;D)-\delta.
\label{eq:recovery}
\end{equation}
Equation~\eqref{eq:recovery} defines recovery independently of the alternative's architecture.

\subsection{Audit Scope}

We consider two distinct mechanisms. For generative recommendation, $m_{\mathrm{gen}}$ is autoregressive LM decoding of item identifiers. For agentic recommendation, $m_{\mathrm{agent}}$ is LM-based control over recommendation tools, covering expert selection and combination of their outputs. The alternatives exclude these \emph{serving-time} mechanisms; they may still use representations created offline by pretrained encoders.

The information declaration is essential to interpreting recovery. $B_{\mathrm{task}}$ specifies the permitted user interactions, item content, and history limits, whereas $B_{\mathrm{repr}}$ records the sources of pretrained representations and which are shared. The protocol $\Pi$ fixes the data splits and evaluation procedure. Matched task inputs do not imply matched pretrained knowledge, model capacity, or optimization and supervision budgets. Consequently, successful recovery supports conditional accuracy attainability; it does not imply that pretrained knowledge is unnecessary, that predictions are identical, or that removing $m$ from the original system would preserve its accuracy.

\subsection{Statistical Recovery Decision}
\label{sec:recovery-statistics}

We quantify recovery uncertainty using paired outcomes. For seed $s$, let $A_s$ and $R_s$ denote the target and alternative; their performance difference is
\begin{equation*}
\Delta_{m,s}=U(R_s;D)-U(A_s;D).
\end{equation*}
We jointly resample user-level outcomes $10{,}000$ times and take the fifth percentile of the bootstrap differences as the one-sided $95\%$ lower confidence bound $\operatorname{LCB}_{95}(\Delta_{m,s})$. We regard recovery at an absolute tolerance $\delta$ as statistically supported for a seed when~\cite{wellek2010equivalence}
\begin{equation}
\operatorname{LCB}_{95}(\Delta_{m,s})>-\delta.
\label{eq:recovery-decision}
\end{equation}
A positive lower bound is stronger evidence: it supports superiority on the primary metric rather than recovery within a positive loss tolerance. For the generative audit, the reported decision must hold for \emph{each} of the three training seeds.

To indicate how sensitive recovery is to the permissible loss, we also evaluate relative tolerances $\eta\in\{1\%,2.5\%,5\%,10\%\}$. The seed-specific plug-in margin is $\delta_s=\eta U(A_s;D)$. Because the stored bootstrap distribution concerns $\Delta_{m,s}$ and treats the estimated target score in $\delta_s$ as fixed, these comparisons are \emph{plug-in sensitivity checks}, not formal one-sided $95\%$ relative non-inferiority tests. Formal relative inference would instead resample the joint statistic $T_{\eta,s}=U(R_s;D)-(1-\eta)U(A_s;D)$.

We summarize the plug-in sensitivity across seeds by
\begin{equation*}
r_{\min}=\max_{s\in\{0,1,2\}}
\frac{\max\!\left(0,-\operatorname{LCB}_{95}(\Delta_{m,s})\right)}{U(A_s;D)}.
\end{equation*}
This quantity is the worst-seed lower-bound shortfall expressed relative to positive target performance; it is \emph{not} an observed mean loss. It gives the infimum plug-in tolerance across the seeds, with equality at the threshold insufficient under the strict decision rule in Eq.~\eqref{eq:recovery-decision}. We additionally assess the effect of correcting the three seed-level tests within each dataset in Section~\ref{sec:robustness}; that correction does not establish paper-wide simultaneous coverage.

The agentic audit instead uses paired accuracy intervals for shared-pool ranking and empirical agreement for budgeted selection; numerical agreement is not a non-inferiority decision.

\section{Constructing Recovery Alternatives}
\label{sec:alternatives}

To construct a recovery alternative, we preserve interaction order, item co-occurrence, and item-text semantics while replacing the serving-time procedure that converts them into recommendations. This tests whether these signals support target-level accuracy without LM generation or tool control, without assuming that the target LM uses the same computations.

We instantiate the idea in two settings. \textbf{LIME-Rec} ranks items by fusing expert scores, rather than decoding item identifiers with an LM. \textbf{DART-Rec} operates on the same experts but replaces LM-based tool control: it fuses all available outputs or selects a fixed subset under a call budget. ``Deterministic'' describes the absence of an LM in serving-time control, not the absence of learned parameters in the fusion model.

\subsection{Shared Recommendation Evidence}
\label{sec:shared-evidence}

Let $\mathcal U$ and $\mathcal V$ be the user and item sets. For user $u$, the chronological history is $H(u)=(i_1,\ldots,i_{T_u})$. With maximum history length $L_{\max}$, we retain the latest $L_u=\min(L_{\max},T_u)$ items as $H_{L_u}(u)=(h_1,\ldots,h_{L_u})$, ordered from oldest to newest. Evaluated users have nonempty histories; history-dependent scores are defined as zero if the history is empty.

\paragraph{Sequential evidence}
To capture recent preference changes, we use SASRec~\cite{Kang2018SASRec}, which provides a candidate score $s_{\mathrm{seq}}(u,i)$ based on the user's interaction sequence.

\paragraph{Collaborative evidence}
For complementary co-occurrence evidence, we use item-based collaborative filtering (ItemCF)~\cite{Sarwar2001ItemCF}. Item pairs within $20$ positions in each training sequence contribute symmetric co-occurrence counts, including repeated positional pairs; we retain up to $100$ neighbors for each source item. Writing $\operatorname{cooc}(a,b)$ for these counts and $N_j$ for item $j$'s training frequency, the score is
\begin{equation*}
s_{\mathrm{cf}}(u,i)=\sum_{t=1}^{L_u}\frac{t}{L_u}
\frac{\operatorname{cooc}(h_t,i)}{\sqrt{N_{h_t}N_i}}.
\end{equation*}
The weight $t/L_u$ emphasizes recent interactions; terms with zero frequency are defined as zero.

\paragraph{Semantic evidence}
To connect items beyond direct collaborative evidence, we encode titles, brands, categories, and descriptions \emph{offline} using the frozen \bgebase{} encoder~\cite{Xiao2024BGE}, obtaining normalized $768$-dimensional item embeddings $e_i$. With recency weights $r_t=\exp[-0.1(L_u-t)]$, semantic evidence is
\begin{equation*}
s_{\mathrm{sem}}(u,i)=\left\langle
\operatorname{norm}\!\left(\frac{\sum_{t=1}^{L_u}r_t e_{h_t}}{\sum_{t=1}^{L_u}r_t}\right),e_i\right\rangle,
\end{equation*}
where $\operatorname{norm}(z)=z/(\lVert z\rVert_2+10^{-8})$. Empty or zero history representations yield zero semantic scores. The frozen text encoder supplies representations but does not decode recommendations at serving time.

The experts' raw scores are not directly comparable. We therefore apply per-user min--max normalization separately to each expert over the full item catalog $\mathcal V$, adding $\epsilon=10^{-8}$ to the denominator. Importantly, normalization takes place \emph{before} any candidate-pool restriction, so a tool's score scale does not change when only a subset of items is revealed.

\subsection{Score Fusion and Ranking}
\label{sec:score-fusion}

LIME-Rec adapts the signal mixture to each user through a lightweight gate using history and expert score distributions. It also discounts observed items while retaining them in the default catalog.

The gate uses a $14$-dimensional feature vector $\phi(u)$: two history features and four score-distribution features for each of the three experts. Let $\mathcal J$ be the full catalog or the shared pool over which these \emph{features} are measured. The history features are $\log(1+L_u)/5$ and $|H_{L_u}(u)\cap\mathcal J|/|\mathcal J|$; repeated history items count once. For each expert, the features are its maximum-minus-mean score, top-two score margin, sample standard deviation (denominator $n-1$ for $n=|\mathcal J|$), and fraction of its top-five items that occur in the history. When $\mathcal J$ is a shared pool, the feature statistics use that pool but the expert scores retain full-catalog normalization.

A linear map $g(u)=W\phi(u)+\mathbf b$, with a $4\times14$ weight matrix $W$, produces three mixture logits and one calibration logit. They determine nonnegative expert weights summing to one and a history penalty:
\begin{equation*}
\mathbf w(u)=\operatorname{softmax}(g_{1:3}(u)),\qquad
\lambda(u)=0.10\sigma(g_4(u)).
\end{equation*}
The score used to rank candidate $i$ is
\begin{equation}
s(u,i)=\sum_x w_x(u)\tilde s_x(u,i)-\lambda(u)\,\mathrm{1}\{i\in H_{L_u}(u)\},
\label{eq:final-score}
\end{equation}
where $x$ indexes the sequential, collaborative, and semantic experts and $\tilde s_x$ denotes their normalized scores. This rule replaces LM item-ID decoding without token generation or LM-based expert choice.

Applied to the complete catalog, Eq.~\eqref{eq:final-score} defines LIME-Rec. Unless stated otherwise, the main model retains at most $20$ history items. For agentic shared-pool ranking, the rule instead ranks the common pool using each expert's normalized scores for \emph{all} pool items, including those outside that expert's own top-$20$ list.

Experts are frozen while the gate is fitted using validation targets and cross-entropy over item scores with a fixed logit scale of $20$. Training histories predict validation items; for testing, the validation item is added to the observed history, while test labels remain excluded from fitting. Shared-pool fusion uses a separate gate fitted on validation targets contained in the candidate pool. The fitted fusion therefore combines evidence without serving-time LM decisions.

\subsection{Tool Selection under Call Budgets}
\label{sec:tool-selection}

DART-Rec isolates LM tool control under two access regimes. With all expert outputs available, it uses the adaptive fusion in Section~\ref{sec:score-fusion}, testing whether an LM is needed to combine shared evidence. Under limited calls, it uses a fixed expert subset and predefined weights, testing LM-based \emph{selection} when evidence acquisition is restricted.

Let $q\in\{1,2\}$ be the call budget and $S$ the selected expert subset, satisfying $1\leq|S|\leq q$. DART-Rec selects the sequential expert for $q=1$ and the sequential plus collaborative experts for $q=2$. It applies the same subset to every user under a given budget. The policies are specified before test evaluation and serve as simple fixed alternatives, not claimed optimal tool choices.

The predefined expert weights are
\[
(\alpha_{\mathrm{seq}},\alpha_{\mathrm{cf}},\alpha_{\mathrm{sem}})=(0.60,0.15,0.25),
\]
with no test-time search or fitting. We renormalize weights within the selected subset:
\begin{equation*}
\bar\alpha_x(S)=\frac{\alpha_x}{\sum_{y\in S}\alpha_y}.
\end{equation*}
Each called expert returns its top-$20$ items and their normalized scores, denoted by $\mathcal T_x(u)$, using the full-catalog score normalization from Section~\ref{sec:shared-evidence}. Only returned items enter the candidate set,
\[
\mathcal C_S(u)=\bigcup_{x\in S}\mathcal T_x(u).
\]
For any $i\in\mathcal C_S(u)$, only experts that returned $i$ contribute to its score:
\begin{equation}
\begin{split}
s_S(u,i)={}&\sum_{x\in S}\bar\alpha_x(S)\tilde s_x(u,i)\,\mathrm{1}\{i\in\mathcal T_x(u)\}\\
&-0.10\,\mathrm{1}\{i\in H_{L_u}(u)\},\quad i\in\mathcal C_S(u).
\end{split}
\label{eq:subset-score}
\end{equation}
Candidates are ranked by Eq.~\eqref{eq:subset-score}. Unlike the all-expert regime, this budgeted rule has fixed weights and a fixed history penalty. Crucially, the LM and fixed selectors use \emph{the same} final ranking rule in the budgeted audit; observed differences can therefore be interpreted as differences in chosen evidence rather than different downstream rankers.

\section{Experiments}

We first test recovery without LM decoding (Audit I), then examine LM tool control with shared evidence or limited calls (Audit II). We finally investigate the signals supporting recovery and its boundaries. Benchmark means describe accuracy, paired comparisons assess recovery or superiority, and interventions and subgroup analyses probe explanations and limits.

\subsection{Experimental Setup}

\paragraph{Datasets and protocol}
The main audits use the 5-core versions of Amazon Reviews 2014 Beauty, Toys, and Sports, in which each retained user and item has at least five interactions~\cite{McAuley2015Amazon}. Yelp~\cite{YelpDataset} tests whether the findings extend beyond these domains. Table~\ref{tab:datasets} gives the dataset sizes and their roles in the study.

\begin{table}[!ht]
\centering
\small
\setlength{\tabcolsep}{3.5pt}
\renewcommand{\arraystretch}{1.05}
\caption{Dataset statistics. Amazon: main audits; Yelp: external-domain boundary.}
\label{tab:datasets}
\begin{tabular*}{\columnwidth}{@{\extracolsep{\fill}}ccccc@{}}
\toprule
Dataset & Users & Items & Interactions & Role \\
\midrule
Beauty & 22,363 & 12,101 & 198,502 & Main \\
Toys & 19,412 & 11,924 & 167,597 & Main \\
Sports & 35,598 & 18,357 & 296,337 & Main \\
Yelp & 30,431 & 20,033 & 316,354 & Boundary \\
\bottomrule
\end{tabular*}
\end{table}

We use leave-two-out evaluation: each user's latest interaction is held out for testing, the second latest for validation, and earlier interactions for training. The validation item is appended to the observed history at test time; on the three Amazon datasets, the held-out test item does not occur in that history. Unless otherwise stated, models rank the complete item catalog, including previously interacted items; we separately test filtering these items in Section~\ref{sec:robustness}.

\paragraph{Metrics}
With one held-out relevant item per user, Recall@K (R@K) is the fraction of users for whom that item appears among the top $K$ recommendations, and equals mean Hit@K. NDCG@K additionally rewards hits placed higher in the ranking. We use R@10 as the primary audit metric and report R@5 and NDCG@5/10 to describe ranking quality more broadly. Where candidate pools are restricted, users whose targets are absent from the pool remain in the denominator.

\paragraph{Run provenance and comparison conditions}
Tables~\ref{tab:main_performance} and~\ref{tab:recovery_audit} use the same reproduced TIGER, GRAM, and LIME-Rec runs and settings; recovery additionally uses paired user-level predictions. Component comparisons use separate runs, whose descriptive means do not enter the recovery tests. Generative recovery uses three training seeds; agentic shared-pool ranking uses paired fixed serving runs. Implementation details are in the appendix.

The experts use SASRec~\cite{Kang2018SASRec}, history-trained ItemCF~\cite{Sarwar2001ItemCF}, and frozen offline BGE representations~\cite{Xiao2024BGE}. SASRec is reused within applicable run groups; fusion and calibration use validation-only fitting. These conditions do not imply equal pretrained knowledge or training budgets.

\subsection{Audit I: Generative Recommendation}

Audit I first establishes LIME-Rec's benchmark competitiveness, then tests recovery against LM targets using paired predictions. A benchmark maximum alone does not establish statistically supported recovery.

\paragraph{Benchmark accuracy}
Table~\ref{tab:main_performance} compares LIME-Rec against six reproduced baselines---TIGER~\cite{Rajput2023TIGER}, IDGenRec~\cite{Tan2024IDGenRec}, HSTU~\cite{Zhai2024HSTU}, LC-Rec~\cite{Zheng2024LCRec}, GRAM~\cite{gram}, and MHL~\cite{Wei2026MHL}---using histories capped at $20$. LIME-Rec has the highest reported three-seed mean on all $12$ dataset--metric combinations. These descriptive maxima motivate the paired audit; bold entries do not denote significance.

\begin{table*}[!t]
\centering
\caption{Benchmark results with histories capped at $20$. Values are three-seed means from reproduced experiments; overlapping models share the same experimental runs and settings as Table~\ref{tab:recovery_audit}. R/N: Recall/NDCG. Bold: column maximum, not statistical significance.}
\label{tab:main_performance}
\small
\setlength{\tabcolsep}{2.35pt}
\renewcommand{\arraystretch}{1.05}
\begin{tabular*}{\textwidth}{@{\extracolsep{\fill}}l*{12}{r}@{}}
\toprule
& \multicolumn{4}{c}{Beauty} & \multicolumn{4}{c}{Toys} & \multicolumn{4}{c}{Sports} \\
\cmidrule(lr){2-5}\cmidrule(lr){6-9}\cmidrule(l){10-13}
Method & R@5 & N@5 & R@10 & N@10 & R@5 & N@5 & R@10 & N@10 & R@5 & N@5 & R@10 & N@10 \\
\midrule
TIGER & 0.0386 & 0.0259 & 0.0584 & 0.0320 & 0.0341 & 0.0217 & 0.0545 & 0.0280 & 0.0215 & 0.0175 & 0.0380 & 0.0226 \\
IDGenRec & 0.0463 & 0.0328 & 0.0665 & 0.0393 & 0.0462 & 0.0323 & 0.0651 & 0.0383 & 0.0273 & 0.0186 & 0.0403 & 0.0228 \\
HSTU & 0.0469 & 0.0314 & 0.0704 & 0.0389 & 0.0433 & 0.0281 & 0.0669 & 0.0357 & 0.0258 & 0.0165 & 0.0414 & 0.0215 \\
LC-Rec & 0.0503 & 0.0352 & 0.0715 & 0.0420 & 0.0543 & 0.0385 & 0.0753 & 0.0453 & 0.0259 & 0.0175 & 0.0384 & 0.0216 \\
GRAM & 0.0638 & 0.0451 & 0.0890 & 0.0532 & 0.0695 & 0.0507 & 0.0956 & 0.0592 & 0.0372 & 0.0256 & 0.0551 & 0.0314 \\
MHL & 0.0574 & 0.0424 & 0.0795 & 0.0495 & 0.0672 & 0.0489 & 0.0903 & 0.0564 & 0.0359 & 0.0249 & 0.0511 & 0.0298 \\
\midrule
\textbf{LIME-Rec} & \textbf{0.0699} & \textbf{0.0484} & \textbf{0.1002} & \textbf{0.0581} & \textbf{0.0802} & \textbf{0.0581} & \textbf{0.1101} & \textbf{0.0677} & \textbf{0.0387} & \textbf{0.0270} & \textbf{0.0565} & \textbf{0.0327} \\
\bottomrule
\end{tabular*}
\end{table*}

\paragraph{Paired recovery against LM targets}
We audit three generative targets: GRAM~\cite{gram}, TIGER~\cite{Rajput2023TIGER}, and LIGER~\cite{liger}. Comparisons use the same splits, task inputs, maximum $20$-item history, and full-catalog ranking without filtering previously interacted items. GRAM, the primary target, retains its original pretrained T5 representation~\cite{Raffel2020T5}. TIGER is reproduced with the LETTER implementation~\cite{Wang2024LETTER}; TIGER and LIGER construct semantic item IDs using the same frozen BGE embeddings as LIME-Rec while retaining their serving-time T5 components, and LIGER also employs dense retrieval. Thus, the latter two targets provide comparisons with a matched upstream embedding source, not an assertion of equal total training resources.

Table~\ref{tab:recovery_audit} reports the reproduced target means and the seed-wise paired-bootstrap decisions defined in Section~\ref{sec:recovery-statistics}. The quantity $r_{\min}$ is the worst-seed \emph{plug-in} tolerance required by the observed confidence bounds, not the mean accuracy gap. The audited objects are these reproduced target instances under a common protocol, not results from other evaluation configurations.

\begin{table}[!ht]
\centering
\small
\setlength{\tabcolsep}{3.5pt}
\renewcommand{\arraystretch}{1.05}
\caption{Generative audit: three-seed mean R@10. Sup.: one-sided $95\%$ lower bound $>0$ on every seed. Sens.\ $5\%$: every seed passes the plug-in sensitivity margin; no formal relative non-inferiority claim. $r_{\min}$: worst-seed infimum tolerance, not mean loss or equivalence.}
\label{tab:recovery_audit}
\begin{tabular*}{\columnwidth}{@{\extracolsep{\fill}}cccccc@{}}
\toprule
LM target & Data & Target & LIME-Rec & $r_{\min}$ & Audit result \\
\midrule
\multirow{3}{*}{GRAM} & Beauty & $0.0890$ & $0.1002$ & $0\%$ & Sup. \\
 & Toys & $0.0956$ & $0.1101$ & $0\%$ & Sup. \\
 & Sports & $0.0551$ & $0.0565$ & $2.8\%$ & Sens.\ $5\%$ \\
\addlinespace
\multirow{3}{*}{TIGER} & Beauty & $0.0584$ & $0.1002$ & $0\%$ & Sup. \\
 & Toys & $0.0545$ & $0.1101$ & $0\%$ & Sup. \\
 & Sports & $0.0380$ & $0.0565$ & $0\%$ & Sup. \\
\addlinespace
\multirow{3}{*}{LIGER} & Beauty & $0.0799$ & $0.1002$ & $0\%$ & Sup. \\
 & Toys & $0.0825$ & $0.1101$ & $0\%$ & Sup. \\
 & Sports & $0.0486$ & $0.0565$ & $0\%$ & Sup. \\
\bottomrule
\end{tabular*}
\end{table}

\paragraph{Interpretation of Audit I}
The evidence is strongest on Beauty and Toys. LIME-Rec is superior to GRAM on every seed for both datasets, with the largest relative R@10 gain, $15.2\%$, occurring on Toys. It is also superior to TIGER and LIGER on all three datasets. Across the nine target--dataset comparisons, eight therefore meet the seed-wise superiority criterion. Because TIGER and LIGER use the same BGE item-embedding source as the alternative, their results show that superiority does not rely solely on changing that source.

The remaining comparison, GRAM--Sports, requires a narrower interpretation. Although the LIME-Rec mean is slightly higher, its paired lower bounds do not establish seed-wise superiority; it instead passes the plug-in $5\%$ sensitivity check, with $r_{\min}=2.8\%$. This is not a formal relative non-inferiority result. Moreover, reproduced TIGER and LIGER scores are below their published values. Semantic-ID construction and inclusion of users whose targets were unseen in training are documented protocol differences, but their individual effects have not been isolated (appendix).

Taken together, Audit I provides strong evidence of attainable target-level R@10 without serving-time LM decoding on the evaluated Amazon protocols, with an explicit statistical qualification for GRAM--Sports. It does not identify how the target LMs internally derive their predictions.

\subsection{Audit II: Agentic Recommendation}

Audit II replaces LM tool control. Following Section~\ref{sec:tool-selection}, we test ranking from shared expert outputs separately from selecting one or two tools, distinguishing evidence combination from acquisition.

\paragraph{Ranking from a shared evidence pool}
We sample $1{,}000$ Beauty users without replacement. For each user, we form a fixed candidate pool from the deduplicated union of all experts' top-$20$ lists, averaging $54.7$ items. All controllers are evaluated on these same pools. Target coverage---the share of users whose held-out item is in the pool---is $18.5\%$, and R@10 includes all $1{,}000$ users rather than conditioning on coverage.

The one-shot controllers receive all three expert outputs upfront, and observe-all controllers retrieve all three before ranking. For these variants and deterministic fusion, each expert provides normalized scores and ranks for \emph{every} pool item, including items outside its own returned top-$20$. By contrast, free tool-use controllers receive only the outputs of tools they call, so these comparisons also involve evidence acquisition. The LM inputs contain opaque item identifiers, expert scores and ranks, and history-membership indicators, without item descriptions. Invalid outputs count as empty rankings after the permitted retries.

The evaluated controllers use Qwen3-4B and Qwen3-8B~\cite{qwen3}, or Granite-3.3-8B-Instruct~\cite{granite33}. Prompt-only controllers receive no recommendation-specific training. Two Qwen3-4B controllers use validation-only LoRA~\cite{Hu2022LoRA}: imitation LoRA learns fusion rankings from $4{,}000$ examples; label LoRA uses $4{,}354$ in-pool validation targets for the first position and fusion rankings thereafter. The separate fusion gate uses in-pool validation targets and four cross-entropy epochs, compared with one LoRA epoch. The differing access, fitting sets, and supervision make this a controlled system comparison, not a test of inherent LM capability under equal training budgets.

\begin{figure}[!ht]
\centering
\includegraphics[width=\columnwidth]{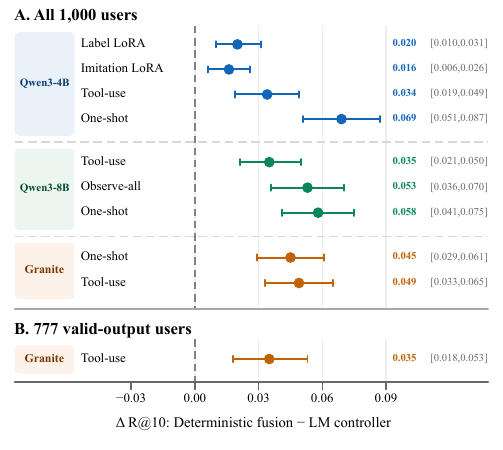}
\caption{Beauty shared-pool R@10 differences (deterministic fusion minus LM controller), with paired two-sided $95\%$ percentile-bootstrap intervals ($10{,}000$ user resamples; unadjusted for multiplicity). A: all $1{,}000$ users; B: Granite tool-use's $777$ valid-output users. Fusion R@10: $0.112$/$0.116$ in A/B. Absolute results are in the appendix.}
\Description{Paired Recall at 10 differences and confidence intervals, grouped by controller. All intervals are above zero. The valid-output Granite comparison uses a separate 777-user subset.}
\label{fig:agentic-ci}
\end{figure}

Figure~\ref{fig:agentic-ci} reports paired R@10 differences between deterministic fusion and the LM controllers. Fusion reaches $0.112$, versus $0.096$ for Qwen3-4B imitation LoRA, an improvement of $16.7\%$ relative to the latter. All nine unadjusted paired $95\%$ intervals in panel A favor fusion. This shows that, with the specified expert information and controller configurations, an LM is not needed to obtain the higher observed shared-pool ranking accuracy.

The advantage is not solely an artifact of invalid LM outputs: fusion also outperforms Granite tool-use on the $777$ users for whom Granite produces valid rankings (panel B). Conversely, the input format limits generalization of the comparison. Supplying item titles, brands, and categories raises Qwen3-8B one-shot R@10 from $0.054$ to $0.075$, although fusion remains at $0.112$. We therefore infer superiority only for the examined tool interfaces, model variants, and supervision conditions, not for unrestricted LM-based ranking.

\paragraph{Selecting tools under a call budget}
Shared-pool ranking holds evidence mostly fixed. Budgeted selection tests a different potential advantage of LM control: deciding which evidence to acquire. We evaluate $1{,}000$ users for each dataset and seed on Beauty, Toys, and Sports. A controller sees the user history before choosing tools, and each called expert reveals its top-$20$ items and scores. Crucially, \emph{all} selectors use the same fixed-weight final ranking rule in Eq.~\eqref{eq:subset-score}; varying the tool selector is therefore the intended comparison.

DART-Rec uses the prespecified fixed policy from Section~\ref{sec:tool-selection}. The LM comparators are a prompt-only Qwen3-4B selector and a validation-only LoRA-adapted selector. The adapted model is supervised by greedy next-tool choices: for state $v$ and candidate tool $t$, the offline utility is $Q(v,t)=\mathrm{1}\{r_t\le10\}/\log_2(r_t+1)$, where $r_t$ is the validation target's rank after adding that tool under fixed-weight ranking. Only states with a uniquely best next tool are used. These labels train immediate tool choice, not multistep planning.

With a single call, LM selectors almost always choose the same sequential expert as the fixed policy, and R@10 differs by no more than $0.001$. The more discriminating budget is two calls. Here, revealed-candidate coverage---the proportion of users whose target appears in the selected tools' returned items---ranges from $8.0\%$ to $14.6\%$ across datasets and seeds. Table~\ref{tab:budgeted} reports mean R@10 and user-level hit agreement under this constraint.

\begin{table}[!ht]
\centering
\small
\caption{Two-call audit: three-seed mean R@10. Exact: identical per-user Hit@10 among six seed--controller comparisons per dataset, not ranking identity or statistical non-inferiority.}
\label{tab:budgeted}
\begin{tabular}{ccccc}
\toprule
Data & Fixed experts & Prompt LM & Adapted LM & Exact \\
\midrule
Beauty & $0.0737$ & $0.0737$ & $0.0737$ & $6/6$ \\
Toys & $0.0833$ & $0.0833$ & $0.0833$ & $6/6$ \\
Sports & $0.0517$ & $0.0520$ & $0.0517$ & $5/6$ \\
\bottomrule
\end{tabular}
\end{table}

The fixed selector and LM selectors yield the same binary Hit@10 outcome for \emph{every user} in $17$ of $18$ seed--controller comparisons; in the remaining comparison, one user differs. Exact top-$10$ ranking agreement is also high: over three seeds, the prompt-only and adapted selectors match fixed-selection rankings for $993.3$--$997.3$ and $995.7$--$1000.0$ of the $1{,}000$ users, respectively, across datasets (Supplementary Table~S13). Nevertheless, binary-hit agreement is not ranking identity and does not constitute a statistical equivalence test. Low candidate coverage and common expert choices may help explain the near-identical accuracy; Eq.~\eqref{eq:recovery-decision} was not applied here.

\paragraph{Interpretation of Audit II}
Shared-pool fusion achieves higher accuracy under the examined inputs and supervision, while fixed selection reproduces almost all LM-controlled hits under limited calls. These complementary findings do not establish the value of LM control for broader planning, alternative tools, or other budgets.

\subsection{What Supports Generative Recovery?}
\label{sec:recovery-evidence}

To examine why LIME-Rec succeeds against GRAM, we compare its components and perturb item--text alignment. The former tests combined signal value; the latter tests meaningful semantic correspondence. Both concern the alternative, not the target LM's internal computation.

\paragraph{Complementary evidence and calibration}
Table~\ref{tab:decomposition} repeats the GRAM recovery decision using canonical paired predictions from LIME-Rec variants. ``Full'' denotes three experts with history calibration; $-$ Calibration removes the history penalty, and $-$ Semantic removes the semantic expert. MiniLM~\cite{Wang2020MiniLM} replaces the BGE offline text encoder. The table reports seed-wise superiority or the first passing plug-in tolerance in the tested grid, keeping its evidential standard aligned with Audit I.

\begin{table}[!ht]
\centering
\caption{GRAM recovery from canonical paired predictions. Sup.: seed-wise lower bound $>0$; NR: no grid margin passes every seed; percentages: first passing margin. $-$ Semantic and Table~\ref{tab:component-comparisons}'s SAS+CF+cal. use different training runs.}
\label{tab:decomposition}
\small
\renewcommand{\arraystretch}{1.05}
\begin{tabular*}{\columnwidth}{@{\extracolsep{\fill}}lccc@{}}
\toprule
Variant & Beauty & Toys & Sports \\
\midrule
Full & \textbf{Sup.} & \textbf{Sup.} & 5\% \\
$-$ Calibration & 5\% & 5\% & NR \\
$-$ Semantic & 2.5\% & 10\% & NR \\
MiniLM & Sup. & Sup. & 10\% \\
\bottomrule
\end{tabular*}
\end{table}

Removing calibration or semantic evidence weakens recovery. On Sports, neither reduced model passes the tested margin grid, whereas Full passes the plug-in $5\%$ criterion. Replacing BGE with MiniLM still supports superiority on Beauty and Toys and passes the plug-in $10\%$ check on Sports. The pattern is consistent with the importance of combining evidence and calibration, without establishing that any particular text encoder is indispensable.

Table~\ref{tab:component-comparisons} asks a related but different question: how much descriptive R@10 is gained as experts and calibration are combined? These are independent component runs under the main experiment's splits, $20$-item history cap, full-catalog ranking, and validation-only fitting protocol. Their means and standard deviations are \emph{not} substituted for the canonical paired predictions in Table~\ref{tab:decomposition}, and the separately run $-$ Semantic and SAS+CF+cal. configurations should not be read as identical statistical trials.

\begin{table}[!ht]
\centering
\caption{Independent component runs under Tables~\ref{tab:main_performance}--\ref{tab:recovery_audit}'s configuration: three-seed R@10 mean $\pm$ sample s.d. These means cannot establish Table~\ref{tab:decomposition}'s paired recovery decisions. SAS/CF/Sem/cal.: SASRec/ItemCF/semantic evidence/history calibration.}
\label{tab:component-comparisons}
\small
\setlength{\tabcolsep}{2pt}
\renewcommand{\arraystretch}{1.05}
\begin{tabular*}{\columnwidth}{@{}l@{\extracolsep{\fill}}ccc@{}}
\toprule
Variant & Beauty & Toys & Sports \\
\midrule
SASRec & $0.0740\!\pm\!0.0014$ & $0.0739\!\pm\!0.0010$ & $0.0395\!\pm\!0.0005$ \\
SAS+cal. & $0.0796\!\pm\!0.0015$ & $0.0773\!\pm\!0.0009$ & $0.0413\!\pm\!0.0010$ \\
SAS+CF+cal. & $0.0912\!\pm\!0.0006$ & $0.0946\!\pm\!0.0004$ & $0.0508\!\pm\!0.0011$ \\
SAS+Sem+cal. & $0.0915\!\pm\!0.0006$ & $0.1001\!\pm\!0.0011$ & $0.0474\!\pm\!0.0007$ \\
Fusion, no cal. & $0.0884\!\pm\!0.0001$ & $0.0982\!\pm\!0.0014$ & $0.0490\!\pm\!0.0005$ \\
\textbf{Full} & $\mathbf{0.1002\!\pm\!0.0009}$ & $\mathbf{0.1101\!\pm\!0.0005}$ & $\mathbf{0.0565\!\pm\!0.0009}$ \\
\bottomrule
\end{tabular*}
\end{table}

The Full configuration exceeds the stronger calibrated two-expert configuration by absolute mean R@10 differences of $0.0087$, $0.0100$, and $0.0057$ on Beauty, Toys, and Sports. Fusion without calibration also improves on calibrated SASRec, while adding calibration increases fusion accuracy on all three datasets. These descriptive patterns corroborate the value of combining signals. Because multiple model changes can accompany a configuration comparison, they do not on their own isolate the causal contribution of the gate.

\paragraph{Does semantic correspondence matter?}
A benefit from adding semantic scores could arise from score-distribution effects rather than meaningful alignment between items and their text. To distinguish these explanations within LIME-Rec, we permute the frozen BGE embeddings across item IDs, holding the seed-$0$ SASRec backbone fixed and refitting the gate and calibration for each of $10$ permutations. The intervention retains an embedding distribution but breaks its assignment to the correct items.

\begin{figure}[!tbp]
\centering
\includegraphics[width=\columnwidth]{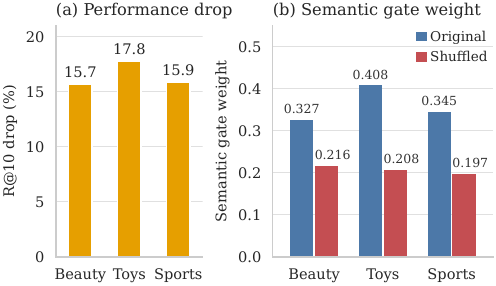}
\caption{Item--text correspondence control. (a) Mean relative R@10 drop over 10 BGE item-embedding permutations. (b) Semantic gate weight in the original system and mean after shuffling. The seed-$0$ SASRec backbone is fixed; gate and calibration are refitted for each permutation.}
\Description{Across Beauty, Toys, and Sports, mean R@10 drops are 15.7, 17.8, and 15.9 percent. Semantic gate weights decrease from 0.327, 0.408, and 0.345 to 0.216, 0.208, and 0.197.}
\label{fig:item-text-correspondence}
\end{figure}

Figure~\ref{fig:item-text-correspondence} shows mean relative R@10 decreases of $15.7\%$, $17.8\%$, and $15.9\%$ on Beauty, Toys, and Sports, respectively, accompanied by lower semantic gate weights in all three datasets. Correct item--text correspondence therefore carries predictive information for the \emph{alternative}, beyond merely supplying an embedding-based score channel. This intervention remains a single-backbone, ten-permutation analysis; it does not establish how the target LM uses semantic information.

\subsection{Where Does Recovery Weaken?}
\label{sec:robustness}

We next stress-test recovery across ranking protocols, domains, and history distributions.

\paragraph{Ranking and statistical robustness}
The default protocol retains previously interacted items in the ranking catalog. If they are instead filtered from \emph{both} systems, fixed-weight fusion still recovers GRAM: it is superior on Beauty and Toys and passes the plug-in $5\%$ sensitivity check on Sports, with $r_{\min}=3.8\%$. The learned-gate variants do not retain recovery under this change. Thus, recovery is not wholly dependent on allowing repeated items, but its robustness is sensitive to the fusion configuration.

The statistical conclusion also depends on the confidence requirement. Bonferroni correction for the three seed-level tests \emph{within each dataset} preserves the reported plug-in grid decisions; the corrected worst-seed GRAM--Sports tolerance is $3.9\%$. This sensitivity check does not correct across all datasets, targets, and exploratory experiments, and does not convert the plug-in relative margin into formal non-inferiority inference.

\paragraph{Domain and user-history boundaries}
Transfer beyond the main domain is not automatic. On Yelp, LIME-Rec obtains mean R@10 of $0.0608$, below GRAM's $0.0687$, with worst-seed plug-in tolerance $17.8\%$---outside the tested grid. Both systems use the same leave-two-out protocol and history limit. This failure to recover with the \emph{tested alternative} does not prove that LM inference is necessary on Yelp; it narrows where this particular recovery construction works.

\begin{figure}[!ht]
\centering
\includegraphics[width=\columnwidth]{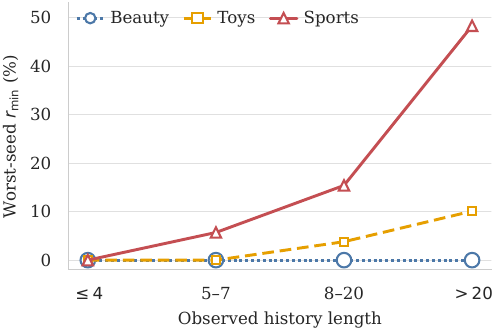}
\caption{Worst-seed plug-in tolerance by pre-truncation history length (training interactions plus validation item). The $>20$ split is post hoc; bounds are descriptive and not multiplicity-adjusted. Subgroup counts are in the appendix; margins follow Section~\ref{sec:recovery-statistics}.}
\Description{Line chart with history bins at most four, five to seven, eight to twenty, and above twenty interactions. Worst-seed recovery tolerances in percent are Beauty: 0, 0, 0, 0; Toys: 0, 0, 3.8, 10.1; Sports: 0, 5.7, 15.4, 48.3. The longest-history split is post hoc.}
\label{fig:history-profile}
% Seed-level values retained from the former history-profile table (seeds 0/1/2).
% <=4: Beauty 0/0/0; Toys 0/0/0; Sports 0/0/0.
% 5--7: Beauty 0/0/0; Toys 0/0/0; Sports 3.7/3.1/5.7.
% 8--20: Beauty 0/0/0; Toys 3.8/0/0; Sports 15.4/11.0/13.7.
% >20: Beauty 0/0/0; Toys 0/0/10.1; Sports 41.4/48.3/44.6.
\end{figure}

Within Amazon, history length exposes further heterogeneity. Figure~\ref{fig:history-profile} groups users by observed \emph{pre-truncation} history length. Beauty recovery is stable; Toys weakens in the longest-history group; and Sports shows the strongest degradation. For Sports, seed-level relative tolerances are $11.0\%$--$15.4\%$ for histories of $8$--$20$ interactions and $41.4\%$--$48.3\%$ above $20$. Degradation is already present below the $20$-item truncation limit, so the history cap cannot by itself explain the pattern. These subgroup findings are descriptive, and the split above $20$ was selected post hoc.

Together, complementary behavioral and aligned semantic signals support recovery in the main audits, while protocol changes, Yelp, and longer histories expose its limits. These boundaries motivate studying when serving-time LM inference provides additional value.

\FloatBarrier

\section{Discussion and Limitations}
The interpretation of our findings depends on the audit conditions. Recovery by an independent alternative does not establish the causal effect of removing LM inference from an existing system, as differences in pretraining, supervision, and optimization remain. The Yelp and user-history results reveal where the tested alternatives achieve weaker recovery, rather than proving LM inference necessary in those settings. Moreover, the agentic audits examine controlled ranking and tool selection, not open-ended planning. Offline text representations remain available, while Recall@10 does not assess conversation, explanations, diversity, or long-term utility.

\section{Conclusion}
Our findings challenge the assumption that strong recommendation accuracy necessarily requires serving-time LM inference. LIME-Rec achieves target-level performance without autoregressive decoding, while DART-Rec improves shared-pool ranking and largely matches LM controllers in top-10 hit outcomes without LM-based tool control. Together, these results demonstrate that the tested accuracy is attainable without the specified inference mechanisms under the evaluated conditions. This shifts the research focus toward when serving-time LM inference provides additional value.

\clearpage
\nobalance
\bibliographystyle{ACM-Reference-Format}
\bibliography{references}
\clearpage
\appendix
\setcounter{table}{0}
\setcounter{figure}{0}
\setcounter{equation}{0}
\renewcommand{\thetable}{A\arabic{table}}
\renewcommand{\thefigure}{A\arabic{figure}}
\renewcommand{\theequation}{A\arabic{equation}}
\section*{Appendix}
% Appendix content included by main.tex.
This appendix provides experimental and reproducibility details for the main paper.
Section~\ref{supp:common} specifies the common protocol and implementation;
Sections~\ref{supp:generative} and~\ref{supp:agentic} expand the generative and agentic audits;
Section~\ref{supp:support} reports supporting evidence and robustness checks;
and Section~\ref{supp:assets} describes the accompanying code package.

\section{Common Protocol and Reproducibility}\label{supp:common}
\subsection{Data, Splits, and Metrics}\label{supp:data}
We use the Amazon Reviews 2014 Beauty, Toys, and Sports datasets~\cite{McAuley2015Amazon}
and Yelp~\cite{YelpDataset}; Table~\ref*{tab:datasets} of the main paper reports their sizes.

Chronological leave-two-out splitting holds out the latest interaction for test and
the preceding one for validation. Test history appends the validation item.
The conversion preserves supplied sequence order using positional timestamps;
this is not a global temporal split.

R@10 measures a held-out target hit in the top ten. R@5 and NDCG@5/10 are secondary
diagnostics; NDCG discounts hits by rank. The full catalog retains previously seen
items unless history exclusion is stated. Amazon test targets never occur in their
visible histories, so exclusion removes no test target.

The canonical generative audit caps every history-dependent input at the most recent
20 interactions: SASRec, ItemCF, semantic pooling, gate features, and calibration.
The shared-pool and limited-tool studies have separate candidate and access protocols,
defined in Sections~\ref{supp:pool} and~\ref{supp:limited}.
Supporting component experiments retain separate complete run groups, with the same $20$-item history cap, in
Section~\ref{supp:support}; their values are not pooled into canonical recovery decisions.

\subsection{Expert and Model Settings}\label{supp:settings}
The three experts provide complementary sequential, collaborative, and semantic
evidence. We instantiate them with SASRec~\cite{Kang2018SASRec},
ItemCF~\cite{Sarwar2001ItemCF}, and frozen \bgebase{} item
representations~\cite{Xiao2024BGE}, respectively. Item text concatenates title,
brand, category, and description where available; embeddings are computed offline
and normalized to unit length. The retained BGE embedding arrays have shape $|\mathcal V|\times768$: $12{,}101\times768$ for Beauty, $11{,}924\times768$ for Toys, and $18{,}357\times768$ for Sports. Rows index items and columns index embedding coordinates; row counts match the main paper's Table~\ref*{tab:datasets}. The main paper defines the expert scoring rules in
Section~\ref*{sec:shared-evidence}.
For each dataset and seed in $\{0,1,2\}$, one independently trained SASRec checkpoint
is reused across the matched configurations that include the sequential expert.
ItemCF counts pairs of distinct items within a 20-position training window,
excludes self-co-occurrences, and retains up to 100 neighbors per source item.
Missing neighbors contribute zero. The semantic scorer divides each vector
by its norm plus $10^{-8}$ and returns zero scores for empty histories
or zero profiles.

\begin{table}[!ht]
\centering\small
\setlength{\tabcolsep}{3pt}\renewcommand{\arraystretch}{1.05}
\caption{Expert architecture and fitting settings.}\label{tab:hyperparameters}
\begin{tabular*}{\columnwidth}{@{\extracolsep{\fill}}p{0.24\columnwidth}p{0.68\columnwidth}@{}}
\toprule
Component & Configuration \\
\midrule
SASRec & Pre-normalized block; hidden size $64$; $2$ blocks; $2$ heads; dropout $0.2$. Canonical maximum history length $20$. \\
SASRec training & Full-vocabulary cross-entropy; Adam; learning rate $10^{-3}$; no weight decay; batch size $128$; up to $200$ epochs; validation every $10$ epochs; patience $30$. \\
ItemCF & Training-only co-occurrence; history window $20$; up to $100$ neighbours per item. \\
Semantic expert & Frozen BGE item embeddings; recency decay $\gamma=0.1$; offline text encoding. \\
Fusion & $14\!\rightarrow\!4$ linear layer; initial weights $(0.60,0.15,0.25)$; score scale $20$; maximum history penalty $0.10$. \\
Full-catalog fusion fitting & Validation only; AdamW; learning rate $0.02$; weight decay $10^{-3}$; batch size $256$; $4$ epochs. \\
Independent repetitions & Full-catalog, repeat-allowed evaluation is shared; all groups cap every history input at $20$ (Section~\ref{supp:support}). Each group retains its own checkpoints and seeds; s.d. is computed within a group. \\
\bottomrule
\end{tabular*}
\end{table}

\subsection{Fusion Gate and Calibration}\label{supp:gate}
This gate architecture defines LIME-Rec on the full catalog and DART-Rec when
all expert outputs are available. For each user and expert $e$, we first min--max
normalize the full-catalog
score vector independently:
\begin{equation}
\widetilde{s}_{e}(u,i)=
\frac{s_{e}(u,i)-\min_j s_{e}(u,j)}
{\max_j s_{e}(u,j)-\min_j s_{e}(u,j)+10^{-8}}.
\end{equation}
The additive constant also defines the behavior for a constant score vector,
which is mapped to zero. The gate input $\phi(u)\in\mathbb{R}^{14}$ uses the capped history $H_{L_u}(u)$ defined in the main paper, with $L_u\le20$. It concatenates
$\log(1+L_u)/5$; four statistics for each normalized expert vector
(top-1 score minus the mean, top-1--top-2 margin, sample standard deviation,
and the fraction of Top-5 items in $H_{L_u}(u)$); and
$|H_{L_u}(u)\cap\mathcal J|/|\mathcal J|$, where $\mathcal J$ is the catalog or shared pool used for the feature statistics. Repeated history items count once in this fraction.
For $n$ scored items, the standard deviation is
$[\sum_{i=1}^{n}(\widetilde s_e(u,i)-\overline s_e(u))^2/(n-1)]^{1/2}$.
Full-catalog features use $n=|\mathcal V|$; shared-pool features use the pool size after full-catalog normalization.

A single linear layer maps $\phi(u)$ to four logits. Its weight matrix is
initialized to zero, its first three biases to $\log(0.60,0.15,0.25)$,
and its fourth bias to $-2$. Softmax over the first
three produces the expert weights, while
$\lambda(u)=0.10\,\mathrm{sigmoid}(z_4)$ produces the bounded history penalty.
The two outputs are trained jointly on validation interactions by minimizing
\begin{equation}
\mathcal{L}=\operatorname{CE}\!\left(
20\left[\sum_e w_e(u)\widetilde{s}_e(u,\cdot)
-\lambda(u)\mathbf{1}[\cdot\in H_{L_u}(u)]\right],y_u\right).
\end{equation}
Validation examples use the training history to predict the validation item;
test examples append the validation item to the training history and predict
the held-out test item. Test interactions never fit or select gate parameters.
For the shuffled-text control, the semantic item-embedding correspondence is
permuted first and the complete gate, including calibration, is then refitted
from scratch on the validation split before test evaluation.

The shared-pool DART-Rec study uses the same architecture with a separate
gate initialized and fitted from scratch, without loading the full-catalog
gate. Fitting uses AdamW, learning rate $0.02$, weight decay $10^{-3}$,
and four epochs, with one variable-sized user pool per update.
Expert scores retain full-catalog
normalization; gate feature statistics and the loss use the candidate pool,
with fitting restricted to validation targets naturally present in that pool.
The limited-tool study instead uses the
fixed subset fusion rule of the main paper, Eq.~(\ref*{eq:subset-score}).

\subsection{Statistical Details}\label{supp:statistics}
The audits use different statistical summaries because they answer different
questions: the generative audit tests recovery relative to a tolerance, whereas
the shared-pool agentic audit compares paired R@10 directly.

For the generative audit, each training seed defines a paired comparison of the
target and LIME-Rec on the same users. Each of $10{,}000$ bootstrap resamples retains
that pairing. The fifth percentile of the R@10 difference is the one-sided
95\% lower confidence bound (LCB); recovery requires this bound to exceed the negative
allowed margin, following Eq.~(\ref*{eq:recovery-decision}) of the main paper.
Relative margins are $1\%,2.5\%,5\%,10\%$ of the seed's estimated target R@10, held fixed when bootstrapping the difference. Positive-margin results are plug-in sensitivity checks, not nominal relative non-inferiority tests; the latter require joint resampling of $T_{\eta,s}$ defined in the main paper. This joint-margin analysis is not available in the retained summaries.
The seed-wise infimum relative tolerance is the nonnegative margin implied by
the LCB, divided by positive target R@10; $r_{\min}$ is its maximum across seeds.
The strict decision inequality need not hold at the infimum itself. The
primary intervals are seed-wise; the within-dataset seed multiplicity sensitivity is reported
separately in Section~\ref{supp:seedresults}.
An all-seed decision requires the criterion on every observed seed.
The seed-level bounds condition on the trained instances; this robustness rule is
not a joint 95\% confidence statement over training-seed populations.

The shared-pool agentic study reports two-sided 95\% paired percentile-bootstrap
intervals for deterministic fusion minus controller R@10. These are the 2.5th and
97.5th percentiles of $10{,}000$ paired user resamples, per comparison and without
multiplicity adjustment. We do not interpret these intervals using generative
non-inferiority margins. Limited-tool comparisons also preserve user pairing;
identical per-user Hit@10 vectors have difference zero and interval $[0,0]$.
All reported bootstrap routines use seed $2027$; the shared-pool table uses a
separate reproducible random stream per comparison.

\section{Audit I: Generative Recommendation}\label{supp:generative}
\subsection{Target Reproduction and Matching Conditions}\label{supp:targets}
GRAM~\cite{gram} is retrained on seeds $\{0,1,2\}$ with its official configuration
and a $20$-item history window. GRAM and LIME-Rec share users, sequences, targets,
and evaluation, while their pretrained representation sources differ (T5 versus BGE).
Implementation and reproduction settings are listed in Section~\ref{supp:reproduction}.

TIGER~\cite{Rajput2023TIGER} is instantiated through the official LETTER
implementation~\cite{Wang2024LETTER}. Its residual-quantization variational
autoencoder (RQ-VAE) turns the same frozen BGE item embeddings used by LIME-Rec
into discrete semantic identifiers, which a pretrained T5 encoder--decoder predicts.
LIGER~\cite{liger} combines dense retrieval with generative scores; its RQ-VAE is
also fitted to those BGE embeddings, and decoding uses beam size $100$.
These controls match pretrained item representations while retaining the target's
T5 components. The shared evaluator includes every test user, including users
whose target item was unseen in training; published LIGER results separate that group.

The retrained TIGER and LIGER scores are below their published values, particularly TIGER on Toys. The BGE semantic-ID source and the shared evaluator are documented protocol differences; the evaluator includes unseen-target users, whereas published LIGER results separate that group. These differences are possible contributors, not a verified decomposition of the reproduction gap. Recovery claims therefore apply to the retained retrained instances.

Table~\ref{tab:target_context} contrasts published and retrained target R@10.
Published values provide context; all recovery decisions use the retrained targets
under the shared evaluator. B/T/S denotes Beauty/Toys/Sports; published LIGER
values use its in-set evaluation with $K=100$.
\begin{table}[H]
\centering\small
\setlength{\tabcolsep}{2.5pt}
\caption{Target-strength context: published and retrained R@10, ordered Beauty/Toys/Sports. Published protocols are not interchangeable with the canonical paired audit; LIGER uses its published in-set evaluation ($K=100$). Retrained columns average three seeds.}\label{tab:target_context}
\begin{tabular*}{\columnwidth}{@{\extracolsep{\fill}}lcc@{}}
\toprule
System & Published B/T/S & Retrained B/T/S \\
\midrule
GRAM & 0.0890/0.0987/0.0554 & 0.0890/0.0956/0.0551 \\
TIGER & 0.0648/0.0712/0.0400 & 0.0584/0.0545/0.0380 \\
LIGER & 0.0845/0.0847/0.0514 & 0.0799/0.0825/0.0486 \\
\bottomrule
\end{tabular*}
\end{table}

\begin{table}[!ht]
\centering\small
\setlength{\tabcolsep}{3pt}
\caption{Information-matching levels in the canonical generative audit. Every row shares splits, task data, a $20$-item history cap, and full-catalog, repeat-allowed evaluation. Training architectures and budgets remain model-specific.}\label{tab:matching-levels}
\begin{tabular*}{\columnwidth}{@{\extracolsep{\fill}}l p{0.28\columnwidth} p{0.36\columnwidth}@{}}
\toprule
System & Upstream item representation & Other pretrained / serving components \\
\midrule
GRAM & Native & T5-based identifier decoding \\
TIGER & Shared frozen BGE & T5-based semantic-ID decoding \\
LIGER & Shared frozen BGE & T5-based decoding and dense retrieval \\
LIME-Rec & Shared frozen BGE & Interaction-trained SASRec, ItemCF, learned fusion; no serving LM \\
\bottomrule
\end{tabular*}
\end{table}
Training targets are the shared observed interactions for each model; the LIME-Rec fusion gate uses validation next-item labels. Equal task inputs and upstream representations do not imply equal trainable parameters or optimization cost. Target fitting settings are listed in Section~\ref{supp:reproduction}.

\subsection{Mean Recovery Results}\label{supp:seedresults}
Table~\ref{tab:recovery_seed_details} expands Table~\ref*{tab:recovery_audit} of
the main paper. Target R@10, LIME-Rec R@10, and $\Delta$ average the three
training seeds. The recovery statistic $r_{\min}$ is the maximum seed-wise
minimum relative tolerance, matching the main paper. The decision column gives
superiority on all seeds, the first passing plug-in grid tolerance on all seeds, or NR
(no common recovery point in the declared grid).

\begin{table}[H]
\centering\small
\setlength{\tabcolsep}{2pt}
\caption{Audit I: three-seed mean R@10 and differences from paired canonical predictions; the native-masked block uses a separate fixed-weight alternative. Sup.: every seed has a positive one-sided $95\%$ lower bound. Sens.@5: all-seed plug-in $5\%$ sensitivity, not formal relative non-inferiority. NR: no grid margin passes. $r_{\min}$ is the maximum seed-wise infimum tolerance.}\label{tab:recovery_seed_details}
\begin{tabular*}{\columnwidth}{@{\extracolsep{\fill}}lccccc@{}}
\toprule
Data & Target & LIME-Rec & $\Delta$ & $r_{\min}$ (\%) & Decision \\
\midrule
\multicolumn{6}{l}{\textit{Primary GRAM audit}} \\
Beauty & 0.0890 & 0.1002 & +0.0112 & 0.0 & Sup. \\
Toys & 0.0956 & 0.1101 & +0.0144 & 0.0 & Sup. \\
Sports & 0.0551 & 0.0565 & +0.0014 & 2.8 & Sens.@5 \\
\midrule
\multicolumn{6}{l}{\textit{Native repeat-masked GRAM}} \\
Beauty & 0.0935 & 0.1025 & +0.0090 & 0.0 & Sup. \\
Toys & 0.1008 & 0.1123 & +0.0115 & 0.0 & Sup. \\
Sports & 0.0579 & 0.0587 & +0.0008 & 3.8 & Sens.@5 \\
\midrule
\multicolumn{6}{l}{\textit{Matched-representation TIGER}} \\
Beauty & 0.0584 & 0.1002 & +0.0418 & 0.0 & Sup. \\
Toys & 0.0545 & 0.1101 & +0.0556 & 0.0 & Sup. \\
Sports & 0.0380 & 0.0565 & +0.0185 & 0.0 & Sup. \\
\midrule
\multicolumn{6}{l}{\textit{Matched-representation LIGER}} \\
Beauty & 0.0799 & 0.1002 & +0.0203 & 0.0 & Sup. \\
Toys & 0.0825 & 0.1101 & +0.0275 & 0.0 & Sup. \\
Sports & 0.0486 & 0.0565 & +0.0079 & 0.0 & Sup. \\
\midrule
\multicolumn{6}{l}{\textit{External-domain Yelp}} \\
Yelp & 0.0687 & 0.0608 & -0.0079 & 17.8 & NR \\
\bottomrule
\end{tabular*}
\end{table}
\FloatBarrier

Table~\ref{tab:primary_lcb} reports three-seed averages. The mean of seed-wise
LCBs is a descriptive summary; recovery decisions retain the seed-level tests.
Differences use unrounded scores. Beauty and Toys have positive lower bounds on
every seed; Sports passes the plug-in $5\%$ grid point on all three seeds.

\begin{table}[H]
\centering\small
\setlength{\tabcolsep}{3pt}\renewcommand{\arraystretch}{1.05}
\caption{Primary GRAM audit: three-seed mean R@10, differences, and seed-level lower bounds.}\label{tab:primary_lcb}
\begin{tabular*}{\columnwidth}{@{\extracolsep{\fill}}lcccc@{}}
\toprule
Data & GRAM & LIME-Rec & $\Delta$ & \shortstack{Mean of\\seed-wise LCBs} \\
\midrule
Beauty & 0.0890 & 0.1002 & +0.0112 & +0.0083 \\
Toys & 0.0956 & 0.1101 & +0.0144 & +0.0111 \\
Sports & 0.0551 & 0.0565 & +0.0014 & -0.0008 \\
\bottomrule
\end{tabular*}
\end{table}

\begin{table}[H]
\centering\small
\setlength{\tabcolsep}{3pt}\renewcommand{\arraystretch}{1.05}
\caption{Secondary metrics for the primary GRAM comparison; three-seed mean $\pm$ s.d.}\label{tab:secondary}
\begin{tabular*}{\columnwidth}{@{\extracolsep{\fill}}llccc@{}}
\toprule
Data & System & R@5 & N@5 & N@10 \\
\midrule
Beauty & GRAM & $0.0638\pm0.0004$ & $0.0451\pm0.0003$ & $0.0532\pm0.0002$ \\
Beauty & LIME-Rec & $0.0699\pm0.0007$ & $0.0484\pm0.0010$ & $0.0581\pm0.0007$ \\
Toys & GRAM & $0.0695\pm0.0004$ & $0.0507\pm0.0001$ & $0.0592\pm0.0004$ \\
Toys & LIME-Rec & $0.0802\pm0.0004$ & $0.0581\pm0.0002$ & $0.0677\pm0.0002$ \\
Sports & GRAM & $0.0372\pm0.0009$ & $0.0256\pm0.0004$ & $0.0314\pm0.0003$ \\
Sports & LIME-Rec & $0.0387\pm0.0004$ & $0.0270\pm0.0002$ & $0.0327\pm0.0003$ \\
\bottomrule
\end{tabular*}
\end{table}

\paragraph{Multiplicity sensitivity.}
Using a Bonferroni adjustment for the three seed-level tests within each dataset
sets the lower-bound percentile to $0.05/3$ rather than $0.05$.
The retained $10{,}000$-resample sensitivity gives mean adjusted seed-level LCBs
$0.0075$, $0.0100$, and $-0.0014$ on Beauty/Toys/Sports, respectively.
Beauty/Toys retain superiority; Sports retains all-seed plug-in sensitivity support at $5\%$ with
worst-seed minimum relative tolerance $3.9\%$. This correction concerns
three seed-level comparisons per dataset, not the complete paper-wide hypothesis family.

\subsection{Candidate-Handling Robustness}\label{supp:candidates}
\paragraph{Native history exclusion.}
GRAM re-runs constrained decoding with the observed capped history removed from
its per-user candidate trie. LIME-Rec removes the same items before ranking.
With fixed expert weights $(0.60,0.15,0.25)$, all Beauty and Toys seeds show superiority
and every Sports seed passes the plug-in sensitivity check at $5\%$ (worst-seed $r_{\min}=3.8\%$).
These native-masking results are included in Table~\ref{tab:recovery_seed_details}.
Refitting the gate under masking instead selects the semantic expert alone
(R@10 $0.0512/0.0641/0.0263$); applying the repeat-allowed-fitted gate to masked
scores selects ItemCF (R@10 $0.0608/0.0773/0.0315$). Neither fitted-routing variant
recovers GRAM. The candidate-handling control therefore reports the fixed-weight
alternative explicitly, rather than assuming fitted-gate stability across conventions.

\paragraph{Post-hoc history filtering.}
A separate sensitivity removes history items from both systems' already-generated
repeat-allowed rankings, with no target re-decoding. GRAM's fixed exported ranking
is filtered, while native masking changes its constrained decoding.
Table~\ref{tab:posthoc_repeat} preserves this distinction and summarizes the three seeds. The mean of seed-wise LCBs is descriptive
and does not replace the per-seed tests.
Beauty/Toys retain superiority; Sports passes the all-seed plug-in sensitivity check at $2.5\%$.

\begin{table}[H]
\centering\small
\setlength{\tabcolsep}{3pt}
\caption{Post-hoc history filtering: three-seed means of differences, seed-level lower bounds, and relative tolerance.}\label{tab:posthoc_repeat}
\begin{tabular*}{\columnwidth}{@{\extracolsep{\fill}}lccc@{}}
\toprule
Data & $\Delta$ & \shortstack{Mean of\\seed-wise LCBs} & Mean tol. (\%) \\
\midrule
Beauty & $+0.0126$ & $+0.0097$ & $0.0$ \\
Toys & $+0.0177$ & $+0.0142$ & $0.0$ \\
Sports & $+0.0025$ & $+0.0004$ & $0.6$ \\
\bottomrule
\end{tabular*}\end{table}

\subsection{External-Domain and History Boundaries}\label{supp:boundaries}
On Yelp, the same leave-two-out, three-seed, $20$-item-history audit compares retrained
GRAM to LIME-Rec. GRAM reaches R@10 $0.0687\pm0.0005$ and LIME-Rec
$0.0608\pm0.0018$. The mean difference is $-0.0079$ and the worst-seed minimum relative tolerance
is $r_{\min}=17.8\%$, outside the declared grid.
Without calibration, LIME-Rec reaches $0.0654\pm0.0017$ with tolerances
$6.1\%$--$10.6\%$; without semantic evidence, $0.0530\pm0.0023$ with
$24.3\%$--$29.7\%$. No tested variant supports one common grid point on all seeds.
These are the boundary results discussed in Section~\ref*{sec:robustness} of the
main paper, rather than entries in its Amazon-only generative table.

For history stratification, $z$ counts interactions visible before the test recommendation.
The original buckets are $z\le4$, $5\le z\le7$, and $z\ge8$; splitting the last bucket
at $20$ is post hoc. Table~\ref{tab:history} provides a complementary three-seed mean profile;
Figure~\ref*{fig:history-profile} of the main paper reports the corresponding
worst-seed $r_{\min}$. These descriptive subgroup
analyses are not multiplicity-adjusted. The Sports gap already occurs at
$8$--$20$ interactions, before truncation is active, so the cap alone does not explain it.

\begin{table}[H]
\centering\small
\setlength{\tabcolsep}{3pt}\renewcommand{\arraystretch}{1.05}
\caption{GRAM history profile: three-seed mean minimum relative tolerance (\%).}\label{tab:history}
\begin{tabular*}{\columnwidth}{@{\extracolsep{\fill}}lcccc@{}}
\toprule
Data & $z\le4$ & $5\le z\le7$ & $8\le z\le20$ & $z>20$ \\
\midrule
Beauty & $0.0$ & $0.0$ & $0.0$ & $0.0$ \\
Toys & $0.0$ & $0.0$ & $1.3$ & $3.4$ \\
Sports & $0.0$ & $4.2$ & $13.4$ & $44.8$ \\
\bottomrule
\end{tabular*}
\end{table}

\begin{table}[!ht]
\centering\small
\caption{History-profile subgroup sizes. History length is measured before the common $20$-item truncation and includes the validation item. Users and bins are shared across seeds.}\label{tab:history-counts}
\begin{tabular*}{\columnwidth}{@{\extracolsep{\fill}}lrrrr@{}}
\toprule
Data & $z\le4$ & $5\le z\le7$ & $8\le z\le20$ & $z>20$ \\
\midrule
Beauty & 7162 & 8712 & 5470 & 1019 \\
Toys & 6579 & 7610 & 4415 & 808 \\
Sports & 11416 & 14209 & 8854 & 1119 \\
\bottomrule
\end{tabular*}
\end{table}
A zero reported $r_{\min}$ is a confidence-bound-derived tolerance, not exact outcome equality. The plotted maximum summarizes seed-wise bounds rather than a confidence interval across subgroup or seed populations.

\FloatBarrier
\section{Audit II: Agentic Recommendation}\label{supp:agentic}
\subsection{Shared-Pool Protocol}\label{supp:pool}
DART-Rec and the LM controllers receive the same users, histories, expert outputs,
and candidate pool. Each expert's top-$20$ list is merged in sequential--collaborative--semantic
order, keeping the first occurrence of an item. Ties are broken lexicographically
by item ID. The resulting pool contains on average $54.7$ items (range $37$--$60$).
Users are sampled uniformly without replacement using NumPy random seed $2027$, then sorted and frozen with the expert cache; one serving run is retained per controller. The held-out target is in the pool for $18.5\%$ of the $1{,}000$ Beauty users.
All users remain in the evaluation denominator; targets outside the pool contribute zero.
DART-Rec retains full-catalog score normalization and computes gate features within the pool and fits its
fusion/calibration layer on validation pools only.

A one-shot controller receives all expert evidence upfront and returns a ranking.
A tool-use controller initially observes user state and obtains expert outputs
through tool calls. Observe-all forces access to all three tools before ranking.
Each queried tool returns item IDs, full-catalog normalized scores, and ranks
for every pool item, including items outside that expert's own top-$20$ list.
Fusion uses all three experts' scores for every pool item. One-shot and
observe-all have the same score-access scope; free tool-use receives only called
experts' outputs. Tools cannot expose targets, correctness feedback, or uncalled
scores. Candidate hashes and expert-score hashes are checked against the
corresponding cache.

The label-adapted condition uses a rebuilt cache. Its recorded equivalence check
preserves the users, candidate pools, and fusion rankings, with maximum expert-score
deviation $3.58\times10^{-7}$ from the original cache. The scores are therefore
not byte-identical across these environments.

We evaluate Qwen3-4B/8B~\cite{qwen3} and Granite-3.3-8B-Instruct~\cite{granite33};
serving settings are recorded in Section~\ref{supp:serving}.

Controllers receive opaque item IDs, history-membership flags, and only expert
outputs permitted by the corresponding access condition. They receive no future
target, correctness feedback, or raw item text. Outputs must follow the prescribed
JSON schema. Exact system prompts are in Section~\ref{supp:prompts}.
A valid finish action contains at least ten unique pool IDs. Unknown tools,
repeated tools, duplicate IDs, out-of-pool IDs, short rankings, and malformed JSON
fail validation. After the allowed retries are exhausted, the response becomes an
empty ranking and contributes zero accuracy; format failures remain in the denominator.
The Granite valid-output comparison uses its $777$ valid users and recomputes
DART-Rec on exactly that subset. It is reported separately from the all-user comparison.

\subsection{Controller Adaptation}\label{supp:adaptation}
Low-rank adaptation (LoRA)~\cite{Hu2022LoRA} adds trainable low-rank parameter updates
to Qwen3-4B. Both adaptations use rank $16$, scaling $32$, no LoRA dropout,
learning rate $10^{-4}$, one epoch, and training seed $2027$.
Imitation adaptation uses $4{,}000$ validation examples and targets the ranking
from validation-fitted deterministic fusion. Label adaptation uses all $4{,}354$
validation users whose held-out item is in the candidate pool (19.5\% of $22{,}363$);
the true item is placed first and the remaining nine entries follow the fusion ranking.
The other $18{,}009$ validation users have no legal pool label and are excluded.
The selected users have mean history length $8.0$ versus $6.6$ outside the subset.
Prompt, pool, serving budget, and test configuration remain frozen; test users
never fit the controllers. Label adaptation supplies a first-position validation label independent of fusion, while the remaining nine target positions use fusion rankings. It therefore differs from pure imitation but is not wholly teacher-independent.

\paragraph{Access and supervision budgets.}
Each method uses one fixed configuration; there is no hyperparameter-search sweep. The fixed sequential--collaborative selector and base weights are specified before test evaluation, and enumeration of other subsets reports diagnostic results rather than selecting the test winner. Table~\ref{tab:controller-budgets} separates supervision and access conditions.
\begin{table}[!ht]
\centering\small
\setlength{\tabcolsep}{2.5pt}
\caption{Declared controller supervision and access. Validation targets are the only fitting labels; optimization budgets are method-specific.}\label{tab:controller-budgets}
\begin{tabular*}{\columnwidth}{@{\extracolsep{\fill}}p{0.25\columnwidth}p{0.30\columnwidth}p{0.32\columnwidth}@{}}
\toprule
Method & Fitting examples / epochs & Test access / trained parameters \\
\midrule
Fusion gate & In-pool validation targets; $4$ epochs & All expert scores; $4\times14+4=60$ parameters \\
Prompt LM & No recommendation fitting & One-shot: all experts; observe-all: all three tools; free tool-use: chosen tools \\
Imitation LoRA & $4{,}000$ validation rankings; $1$ epoch & All expert evidence; rank-$16$ updates to Qwen3-4B \\
Label LoRA & $4{,}354$ in-pool validation labels; $1$ epoch & All expert evidence; rank-$16$ updates to Qwen3-4B \\
Fixed selector & No selector fitting & History before selection; at most $q$ experts; no selector parameters \\
Adapted selector & $10{,}753$--$13{,}226$ validation states per dataset--seed; $1$ epoch & History and called outputs; rank-$16$ updates \\
\bottomrule
\end{tabular*}
\end{table}
The shared-pool gate and label LoRA use validation targets naturally present in the same expert-pool protocol, whereas imitation LoRA learns a subset of fusion rankings. LoRA supervision and gate cross-entropy therefore differ. Access-matched one-shot and observe-all comparisons assess ranking; free tool-use also assesses evidence acquisition. Prompt formats and retry limits are specified in Sections~\ref{supp:prompts} and~\ref{supp:serving}.

\subsection{Limited-Tool Protocol}\label{supp:limited}
This study uses $1{,}000$ users per Amazon dataset--seed pair and budgets $q=1,2$.
Before the first call, a controller sees history only. Each distinct called expert
returns its top-$20$ items and full-catalog normalized scores; the initial view has
no candidate union or uncalled evidence. Every selector uses the same fixed final
fusion rule in Eq.~(\ref*{eq:subset-score}) of the main paper: renormalized
$(0.60,0.15,0.25)$ weights over called experts, zero contribution for absent returned
items, and a $0.10$ history penalty. The base weights were manually specified
before test evaluation, without selection using test-set performance.
The sequential policy at $q=1$ and the
sequential--collaborative pair at $q=2$ were specified and frozen before test evaluation.
The pair retains sequential and collaborative behavioral evidence within
the two-call budget.
The evaluator also enumerates the remaining fixed combinations; a three-call run
is a separate reference rather than a same-budget comparator.

The prompt-only and adapted selectors use Qwen3-4B. Selector LoRA uses greedy next-step NDCG@10 supervision from held-out validation targets. For state $v$ and available tool $t$, let $r_t$ be the 1-based target rank after appending that tool's cached output and applying the fixed subset fusion rule. The utility is
\[
Q(v,t)=\begin{cases}1/\log_2(r_t+1),&r_t\le10,\\0,&\text{otherwise}.\end{cases}
\]
The label is $\arg\max_t Q(v,t)$ only when the maximum is unique (exact numerical equality defines ties). This is a one-step oracle label, not a look-ahead policy over the remaining budget. The offline label generator accesses all experts' validation outputs and the validation target; neither uncalled outputs nor the target appear in the controller input. Test targets never generate labels.

Each validation user generates five states: the empty prefix for $q=1$, and the empty prefix plus each one-tool prefix for $q=2$. All-zero and other tied maxima are excluded. The retained manifests report $10{,}753$--$13{,}226$ examples per dataset--seed. Averaged over three seeds, the excluded fractions are $88.29\%$, $88.81\%$, and $93.13\%$ for Beauty, Toys, and Sports. Consequently, adaptation learns from the subset with a uniquely informative next action, not the entire state distribution. Retained examples use the same public messages as evaluation.
The adaptation configuration is rank $16$, scaling $32$, one epoch, and learning
rate $10^{-4}$. The exact system prompt is in Section~\ref{supp:prompts}.
The user message serializes history, called tools, available tools, remaining
budget, and observed results as JSON. A response must contain exactly one string
\texttt{tool} field naming an available expert. Unknown/repeated tools and malformed
JSON produce an empty ranking without fallback. Transport or serving errors
interrupt the run rather than count as completed results. Each successful controller
uses the specified budget; output generation is capped at $128$ tokens and
Qwen thinking is disabled.

\subsection{Full Results}\label{supp:agentresults}
\paragraph{Shared-pool controller comparison.}
\begin{table}[t]
\centering
\caption{Beauty shared-pool audit: all $1{,}000$ users (A) and Granite tool-use's $777$ valid-output users (B). $\Delta$: paired R@10 difference, deterministic fusion minus the LM controller.}
\label{tab:agentic-full}
\small
\setlength{\tabcolsep}{4.2pt}
\renewcommand{\arraystretch}{1.05}
\begin{tabular}{@{}ccccc@{}}
\toprule
Model & Setting & R@10 & $\Delta$ & 95\% CI \\
\midrule
\multicolumn{5}{@{}c}{\textit{Panel A: All 1,000 users}} \\[1pt]
\textbf{Deterministic fusion} & --- & $\mathbf{0.112}$ & --- & --- \\
\addlinespace
\multirow{4}{*}{Qwen3-4B} & Label / 1-shot & $0.092$ & $+0.020$ & $[0.010,0.031]$ \\
 & Imit. / 1-shot & $0.096$ & $+0.016$ & $[0.006,0.026]$ \\
 & Tools & $0.078$ & $+0.034$ & $[0.019,0.049]$ \\
 & 1-shot & $0.043$ & $+0.069$ & $[0.051,0.087]$ \\
\addlinespace
\multirow{3}{*}{Qwen3-8B} & Tools & $0.077$ & $+0.035$ & $[0.021,0.050]$ \\
 & Observe-all & $0.059$ & $+0.053$ & $[0.036,0.070]$ \\
 & 1-shot & $0.054$ & $+0.058$ & $[0.041,0.075]$ \\
\addlinespace
\multirow{2}{*}{Granite} & 1-shot & $0.067$ & $+0.045$ & $[0.029,0.061]$ \\
 & Tools & $0.063$ & $+0.049$ & $[0.033,0.065]$ \\
\midrule
\multicolumn{5}{@{}c}{\textit{Panel B: Valid outputs (777 users)}} \\[1pt]
\textbf{Deterministic fusion} & --- & $\mathbf{0.116}$ & --- & --- \\
Granite & Valid / Tools & $0.081$ & $+0.035$ & $[0.018,0.053]$ \\
\bottomrule
\end{tabular}

\par\addvspace{2pt}
\parbox{\columnwidth}{\footnotesize
\textit{Notes.} Label and Imit. denote label-supervised and imitation LoRA, respectively. Tools denotes tool-use access. Both panels report two-sided $95\%$ paired percentile-bootstrap confidence intervals for $\Delta$ ($10{,}000$ user-level resamples). The intervals are per-comparison and are not multiplicity-adjusted.
}
\end{table}

The shared-pool comparisons use the two-sided intervals and sign convention
of Figure~\ref*{fig:agentic-ci} of the main paper; Table~\ref{tab:agentic-full} retains the absolute results. All nine all-user intervals are
above zero. Imitation adaptation increases Qwen3-4B one-shot R@10 from $0.043$ to
$0.096$ ($+0.053$, paired 95\% CI $[0.037,0.070]$); DART-Rec remains higher by
$0.016$ ($[0.006,0.026]$).

On the Granite valid-output subset, DART-Rec reaches R@10 $0.116$ versus $0.081$;
the difference is $+0.035$ with CI $[0.018,0.053]$. The tool-use format-failure rate is $22.3\%$.

Additional shared-pool controls retain the same pool. Removing semantic-tool access
yields Qwen3-8B/4B/Granite R@10 $0.077/0.077/0.065$ (format failures
$0.001/0.009/0.214$). Observe-all reaches $0.059/0.062$ for Qwen3-8B/4B
(format failures $0.019/0.057$). Exposing title, brand, and category, but not description,
yields Qwen3-8B tool-use/one-shot R@10 $0.078/0.075$, compared with opaque-ID
$0.077/0.054$. The metadata-visible controller-minus-fusion differences are
$-0.034$ with CI $[-0.048,-0.019]$ and $-0.037$ with CI $[-0.051,-0.023]$.

\paragraph{Limited-tool comparison.}
Table~\ref{tab:limited_seeds} averages the three seeds for each limited-tool dataset at $q=2$.
For each controller, $\Delta$ is fixed sequential--collaborative R@10 minus controller
R@10. Seventeen paired Hit@10 differences are zero with percentile-bootstrap CI $[0,0]$; these degenerate intervals reflect identical observed outcomes and are not population-equivalence certificates. No margin-based decision from the main paper Eq.~\eqref{eq:recovery-decision} is asserted for this study; the prompt-only
Sports seed-$1$ difference is $-0.001$ with CI $[-0.003,0]$, corresponding to one user.

\begin{table}[H]
\centering\small
\setlength{\tabcolsep}{3pt}\renewcommand{\arraystretch}{1.05}
\caption{Limited-tool audit ($q=2$): three-seed mean R@10 and paired differences.}\label{tab:limited_seeds}
\begin{tabular*}{\columnwidth}{@{\extracolsep{\fill}}lccccc@{}}
\toprule
Data & Fixed & Prompt & Adapted & $\Delta_{\mathrm{prompt}}$ & $\Delta_{\mathrm{adapted}}$ \\
\midrule
Beauty & $0.0737$ & $0.0737$ & $0.0737$ & $+0.0000$ & $+0.0000$ \\
Toys & $0.0833$ & $0.0833$ & $0.0833$ & $+0.0000$ & $+0.0000$ \\
Sports & $0.0517$ & $0.0520$ & $0.0517$ & $-0.0003$ & $+0.0000$ \\
\bottomrule
\end{tabular*}
\end{table}

\begin{table}[H]
\centering\small
\setlength{\tabcolsep}{3pt}\renewcommand{\arraystretch}{1.05}
\caption{Mean users with exactly matching rankings versus the fixed pair, out of $1{,}000$; averaged over three seeds.}\label{tab:rankagreement}
\begin{tabular*}{\columnwidth}{@{\extracolsep{\fill}}lcc@{}}
\toprule
Data & Prompt & Adapted \\
\midrule
Beauty & $993.3$ & $999.3$ \\
Toys & $996.7$ & $995.7$ \\
Sports & $997.3$ & $1000.0$ \\
\bottomrule
\end{tabular*}
\end{table}

Exact ranking agreement (Table~\ref{tab:rankagreement}) differs from identical
Hit@10 outcomes. Adaptation gives exact rankings on five of nine dataset--seed pairs;
its per-user Hit@10 and NDCG@10 match the fixed pair for at least $998$ users in
every comparison. At $q=1$, controllers use the sequential expert for nearly all users,
with maximum aggregate R@10 difference $0.001$.
\paragraph{Serving diagnostics.}
At $q=2$, successful LM runs use exactly $2.0$ calls/user, format-failure rate $0$,
and approximately $1{,}190$ input/$17$ output tokens/user; fixed selection uses no LM tokens. Cache-replay timing excludes expert inference.
The shared-pool deployment records $3.1$ ms for CPU fusion and $3.07$--$4.45$ s
for resident-GPU 8B controllers; these different-hardware measurements are not speedup ratios.

\section{Supporting Evidence and Robustness}\label{supp:support}
These analyses are supporting or diagnostic comparisons, not inputs to the
primary recovery decisions. For TIGER, GRAM, and LIME-Rec, the benchmark and canonical audit use the same experimental runs and evaluation settings. Component, expert, and robustness analyses retain separate run groups. All use full-catalog, repeat-allowed evaluation and cap all history inputs at $20$. Table~\ref{tab:history_protocols} identifies the
run groups and their roles. Each group is reported intact, with seed variability
summarized within that group.
\begin{table}[H]
\centering\small
\setlength{\tabcolsep}{3pt}
\caption{Evaluation sources and scientific roles; benchmark and main audit share runs for TIGER, GRAM, and LIME-Rec.}\label{tab:history_protocols}
\begin{tabular}{@{}p{0.31\columnwidth}p{0.41\columnwidth}p{0.20\columnwidth}@{}}
\toprule
Experiment group & Reported outputs & Role \\
\midrule
Benchmark LIME-Rec (shared audit runs) & Main Table~\ref*{tab:main_performance} & Full-metric comparison \\
Main audit & Main Tables~\ref*{tab:recovery_audit} and \ref*{tab:decomposition} & Paired recovery \\
Component repetitions & Main Table~\ref*{tab:component-comparisons}; Table~\ref{tab:components} & Factorial effects \\
Expert and robustness repetitions & Tables~\ref{tab:ablation}--\ref{tab:weakbackbone}; Figure~\ref{fig:expert-combinations} & Supporting evidence \\
\bottomrule
\end{tabular}
\end{table}

\paragraph{Benchmark and audit run groups.}
All experiments, including TIGER, IDGenRec, HSTU, LC-Rec, GRAM, and MHL in main Table~\ref*{tab:main_performance}, are complete reproductions. All use a $20$-item history limit. For TIGER, GRAM, and LIME-Rec, main Tables~\ref*{tab:main_performance} and~\ref*{tab:recovery_audit} use the same experimental runs and evaluation settings. LIME-Rec and the component group also cap semantic pooling, gate features, and calibration at $20$. The component analyses use separate runs and are not combined with the paired audit for statistical inference. The benchmark reports descriptive means, while the audit additionally uses paired user-level predictions to assess seed-wise superiority and plug-in recovery sensitivity.

\subsection{Component Contributions}\label{supp:components}
Table~\ref{tab:components} measures three-seed component effects in one complete
repetition group, whereas Table~\ref{tab:ablation} reports seed-$0$ expert subsets
from the expert-composition group. Their full-model estimates differ across
independent runs. All history-dependent inputs, including semantic pooling, gate features, and calibration, are capped at the most recent $20$ interactions in the component and canonical audit groups. Each reported sample standard
deviation describes variation over seeds $0,1,2$ within its group, not variation
across experimenters or a pooled collection of runs. Formal recovery is assessed
with the paired predictions from the main audit group; aggregate means from
other groups are not substituted into those tests.

\begin{table}[H]
\centering\small
\setlength{\tabcolsep}{2pt}
\caption{Independent component repetitions: R@10 mean $\pm$ sample s.d. over three seeds, using the same configuration as main Tables~\ref*{tab:main_performance} and~\ref*{tab:recovery_audit}. These means cannot establish the canonical paired recovery decisions in main Table~\ref*{tab:decomposition}.}
\label{tab:components}
\begin{tabular*}{\columnwidth}{@{}l@{\extracolsep{\fill}}ccc@{}}
\toprule
Variant & Beauty & Toys & Sports \\
\midrule
SASRec & $0.0740\!\pm\!0.0014$ & $0.0739\!\pm\!0.0010$ & $0.0395\!\pm\!0.0005$ \\
SAS+cal. & $0.0796\!\pm\!0.0015$ & $0.0773\!\pm\!0.0009$ & $0.0413\!\pm\!0.0010$ \\
SAS+CF+cal. & $0.0912\!\pm\!0.0006$ & $0.0946\!\pm\!0.0004$ & $0.0508\!\pm\!0.0011$ \\
SAS+Sem+cal. & $0.0915\!\pm\!0.0006$ & $0.1001\!\pm\!0.0011$ & $0.0474\!\pm\!0.0007$ \\
Fusion, no cal. & $0.0884\!\pm\!0.0001$ & $0.0982\!\pm\!0.0014$ & $0.0490\!\pm\!0.0005$ \\
Full model & $0.1002\!\pm\!0.0009$ & $0.1101\!\pm\!0.0005$ & $0.0565\!\pm\!0.0009$ \\
\bottomrule
\end{tabular*}
\end{table}
SAS, CF, Sem, and cal. denote SASRec, ItemCF, semantic evidence, and calibration.
The full model adds $0.0087$/\allowbreak$0.0100$/\allowbreak$0.0057$ above the
higher-R@10 calibrated pair on Beauty/Toys/Sports (displayed means). These aggregate diagnostic
differences are separate from the paired canonical ablations in
Table~\ref*{tab:decomposition} of the main paper.

\begin{figure*}[t]
\centering
\includegraphics[width=\textwidth]{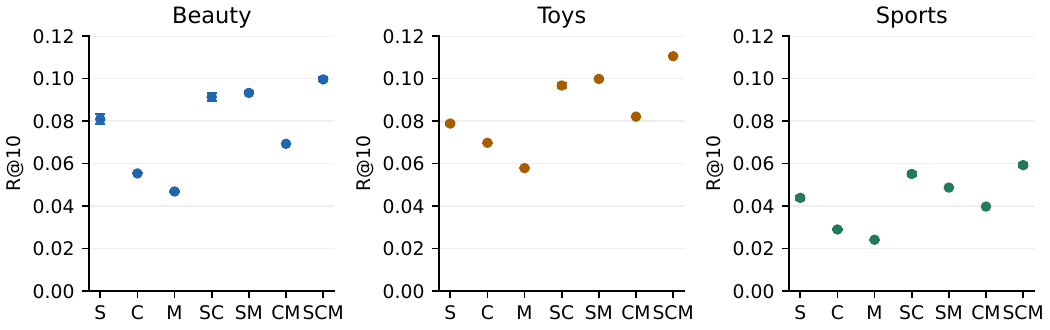}
\caption{Expert complementarity across independent repetitions under the full-catalog, repeat-allowed protocol. S: SASRec; C: ItemCF; M: semantic expert; concatenated letters denote subsets (SCM: all three). Points show R@10 means over seeds $0,1,2$; error bars show sample standard deviations. Every subset retains validation-fitted bounded history calibration. These repetitions support component analysis; recovery uses the main audit's paired predictions.}
\Description{Three panels compare the seven nonempty subsets of sequential, collaborative, and semantic experts on Beauty, Toys, and Sports. All panels show three-seed means and sample standard deviations; the full three-expert combination has the highest mean R@10 in each panel.}
\label{fig:expert-combinations}
\end{figure*}

The seed-$0$ expert-composition results in Table~\ref{tab:ablation} retain bounded
history calibration in every row. The fusion--calibration factors are isolated in the main paper's
Table~\ref*{tab:component-comparisons} in the component repetition group.
Figure~\ref{fig:expert-combinations} summarizes all seven expert subsets over
three seeds in the expert-composition repetition group. Adding semantic evidence to SASRec
improves mean R@10 on all three datasets, and the full combination has the
highest mean. These comparisons retain bounded calibration for every subset.

\begin{table}[H]
\centering\small
\setlength{\tabcolsep}{2.5pt}
\caption{Seed-$0$ expert-composition diagnostics; R is Recall and N is NDCG.}\label{tab:ablation}
\begin{tabular*}{\columnwidth}{@{\extracolsep{\fill}}lcccc@{}}
\toprule
Configuration & R@5 & N@5 & R@10 & N@10 \\
\midrule
\multicolumn{5}{l}{\textit{Beauty}} \\
SASRec only & 0.0582 & 0.0414 & 0.0794 & 0.0483 \\
ItemCF only & 0.0375 & 0.0251 & 0.0554 & 0.0308 \\
Semantic only & 0.0297 & 0.0199 & 0.0469 & 0.0254 \\
SASRec + ItemCF & 0.0650 & 0.0466 & 0.0896 & 0.0545 \\
SASRec + Semantic & 0.0647 & 0.0447 & 0.0926 & 0.0536 \\
ItemCF + Semantic & 0.0449 & 0.0290 & 0.0693 & 0.0369 \\
\textbf{Three experts} & \textbf{0.0692} & \textbf{0.0491} & \textbf{0.0986} & \textbf{0.0586} \\
\midrule
\multicolumn{5}{l}{\textit{Toys}} \\
SASRec only & 0.0584 & 0.0420 & 0.0795 & 0.0489 \\
ItemCF only & 0.0495 & 0.0330 & 0.0698 & 0.0395 \\
Semantic only & 0.0377 & 0.0243 & 0.0579 & 0.0307 \\
SASRec + ItemCF & 0.0719 & 0.0529 & 0.0977 & 0.0613 \\
SASRec + Semantic & 0.0704 & 0.0494 & 0.1000 & 0.0590 \\
ItemCF + Semantic & 0.0549 & 0.0380 & 0.0821 & 0.0468 \\
\textbf{Three experts} & \textbf{0.0794} & \textbf{0.0566} & \textbf{0.1106} & \textbf{0.0666} \\
\midrule
\multicolumn{5}{l}{\textit{Sports}} \\
SASRec only & 0.0286 & 0.0194 & 0.0429 & 0.0240 \\
ItemCF only & 0.0191 & 0.0136 & 0.0290 & 0.0167 \\
Semantic only & 0.0147 & 0.0098 & 0.0241 & 0.0128 \\
SASRec + ItemCF & 0.0388 & 0.0269 & 0.0555 & 0.0323 \\
SASRec + Semantic & 0.0314 & 0.0207 & 0.0487 & 0.0262 \\
ItemCF + Semantic & 0.0255 & 0.0176 & 0.0397 & 0.0222 \\
\textbf{Three experts} & \textbf{0.0412} & \textbf{0.0285} & \textbf{0.0599} & \textbf{0.0345} \\
\bottomrule
\end{tabular*}
\end{table}

\subsection{Item--Text Correspondence}\label{supp:shuffle}
\begin{figure}[!ht]
\centering
\includegraphics[width=\columnwidth]{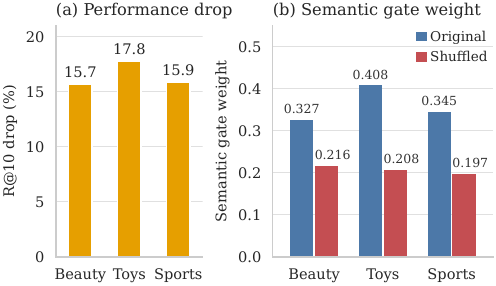}
\caption{Item--text correspondence diagnostic: relative R@10 drop and semantic gate weight averaged over $10$ permutations. The seed-$0$ SASRec backbone is fixed across permutations.}
\Description{Beauty, Toys, and Sports show mean R@10 drops of 15.7, 17.8, and 15.9 percent. Semantic weights change from 0.327, 0.408, and 0.345 to 0.216, 0.208, and 0.197, respectively.}
\label{fig:shuffle}
\end{figure}

The shuffle control permutes BGE embeddings across item IDs, then refits the
complete gate and calibration on validation for each of ten permutations.
The SASRec seed-$0$ backbone is fixed across all permutations under the same evaluation protocol.
R@10 declines by $15.7\%\pm1.0$, $17.8\%\pm0.9$, and $15.9\%\pm0.8$ on
Beauty/Toys/Sports. Mean semantic gate weight changes from $0.327/0.408/0.345$
to $0.216/0.208/0.197$. Figure~\ref{fig:shuffle} depicts these correspondence diagnostics. This
intervention measures predictive dependence on item--text alignment in the
alternative, rather than the internal computation of the audited LM.

\subsection{Expert Non-Redundancy}\label{supp:nonredundancy}
\begin{figure}[!ht]
\centering
\includegraphics[width=0.70\columnwidth]{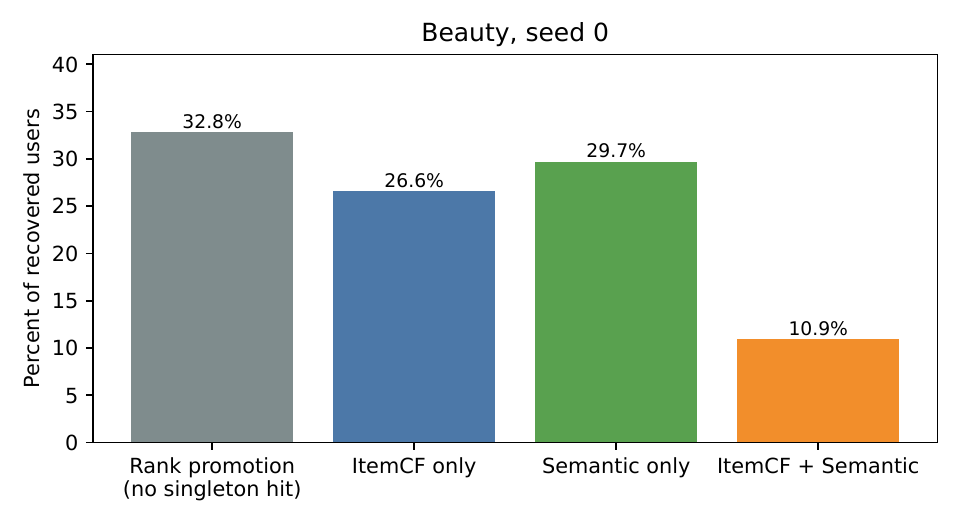}
\caption{Recovered-user decomposition on Beauty seed $0$.}
\Description{The recovered targets divide into rank promotion, collaborative-only, semantic-only, and joint collaborative--semantic singleton hits.}
\label{fig:recovered}
\end{figure}

The target-hit decomposition in Table~\ref{tab:hitdecomp} partitions users according
to which singleton experts retrieve the held-out item in their top-$10$ list.
NONE means no singleton hit, ALL means all three, and a pair column means exactly
that pair hits. The overlap statistics in Table~\ref{tab:mechanism} compare expert
lists using Jaccard@10 and full-catalog average-rank Spearman correlation.
Both use seed-$0$ outputs and $2{,}000$ sampled test users (seed $42$).
\begin{table}[!ht]
\centering\small
\setlength{\tabcolsep}{3pt}
\caption{Exactly which singleton experts hit the target: percentages of sampled test users.}\label{tab:hitdecomp}
\begin{tabular*}{\columnwidth}{@{\extracolsep{\fill}}lrrr@{}}
\toprule
Hit subset & Beauty & Toys & Sports \\
\midrule
NONE & 87.45 & 86.55 & 93.10 \\
only-SAS & 4.35 & 4.20 & 2.95 \\
only-CF & 1.65 & 2.05 & 1.25 \\
only-Sem & 2.65 & 2.55 & 1.40 \\
SAS+CF & 2.30 & 1.95 & 0.75 \\
SAS+Sem & 0.25 & 0.30 & 0.05 \\
CF+Sem & 0.40 & 0.60 & 0.35 \\
ALL & 0.95 & 1.80 & 0.15 \\
\bottomrule\end{tabular*}\end{table}\begin{table}[!ht]
\centering\small
\setlength{\tabcolsep}{3pt}
\caption{Expert non-redundancy; S, C, and M denote SASRec, ItemCF, and semantic evidence.}\label{tab:mechanism}
\begin{tabular*}{\columnwidth}{@{\extracolsep{\fill}}lccc@{}}
\toprule
Data & Pair & Jaccard@10 & Spearman \\
\midrule
Beauty & S--C & 0.0572 & $+0.1065$ \\
Beauty & S--M & 0.0401 & $+0.1092$ \\
Beauty & C--M & 0.0635 & $+0.0816$ \\
Toys & S--C & 0.0524 & $+0.0947$ \\
Toys & S--M & 0.0484 & $+0.1430$ \\
Toys & C--M & 0.0812 & $+0.0799$ \\
Sports & S--C & 0.0191 & $+0.1078$ \\
Sports & S--M & 0.0249 & $+0.2382$ \\
Sports & C--M & 0.0472 & $+0.0938$ \\
\bottomrule\end{tabular*}\end{table}
For Beauty seed $0$, fusion retrieves $64$ targets absent from SASRec's top-$10$:
$21$ ($32.8\%$) are rank-promotion cases where no singleton hits; among all $64$,
$26.6\%$ are collaborative-only hits, $29.7\%$ semantic-only, and $10.9\%$ joint
collaborative--semantic hits. Figure~\ref{fig:recovered} shows this partition.

\subsection{Encoder and Calibration Robustness}\label{supp:encoders}
Table~\ref{tab:robustness} separates calibration/encoder controls (A) from a same-group sequential-expert reference (B). Global calibration replaces $\lambda(u)$ with one shared scalar $\lambda=0.10\,\sigma(a)$, with $a$ initialized to $-2$ and fitted on validation using the same loss. The three expert weights remain user-adaptive; only the penalty loses its dependence on $u$. MiniLM and BGE embeddings~\cite{Wang2020MiniLM,Xiao2024BGE} use matched SASRec checkpoints. The global scalar changes R@10 by at most $0.0007$;
encoder replacement changes it by approximately $0.0011$ at most.
These independent robustness repetitions complement the paired encoder
ablation in the main paper's Table~\ref*{tab:decomposition}.
\begin{table}[H]
\centering\small
\setlength{\tabcolsep}{3pt}
\caption{Independent robustness group: three-seed mean R@10. A: per-user versus one shared validation-fitted history penalty, and MiniLM versus BGE with matched SASRec. B: LIME-Rec versus its SASRec expert in this same group; absolute gains are descriptive, not canonical paired recovery decisions.}\label{tab:robustness}
\begin{tabular*}{\columnwidth}{@{\extracolsep{\fill}}lcccc@{}}
\toprule
\multicolumn{5}{l}{\textit{A: Calibration and encoder controls}} \\
Data & Per-user & Global & MiniLM & BGE \\
\midrule
Beauty & 0.0996 & 0.0990 & 0.0996 & 0.0996 \\
Toys & 0.1105 & 0.1099 & 0.1095 & 0.1105 \\
Sports & 0.0593 & 0.0590 & 0.0584 & 0.0593 \\
\bottomrule\end{tabular*}
\par\addvspace{3pt}
\begin{tabular*}{\columnwidth}{@{\extracolsep{\fill}}lccc@{}}
\toprule
\multicolumn{4}{l}{\textit{B: Sequential-expert reference}} \\
Data & SASRec & LIME-Rec & Abs. gain \\
\midrule
Beauty & 0.0761 & 0.0996 & +0.0235 \\
Toys & 0.0746 & 0.1105 & +0.0359 \\
Sports & 0.0417 & 0.0593 & +0.0176 \\
\bottomrule\end{tabular*}\end{table}
\subsection{Additional SASRec Control}\label{supp:sasrec}
Table~\ref{tab:weakbackbone} substitutes the original one-negative binary
cross-entropy (BCE) and ICDM'18-style SASRec block for full-vocabulary
cross-entropy and the pre-normalized block. Fusion increases R@10 by
$72.9\%/81.4\%/127.2\%$ on Beauty/Toys/Sports (unrounded means).
This matched-backbone diagnostic measures fusion gains; it is not a separate
controlled generative-target recovery test.
\begin{table}[H]
\centering
\small
\setlength{\tabcolsep}{3pt}
\renewcommand{\arraystretch}{1.05}
\begin{tabular*}{\columnwidth}{@{\extracolsep{\fill}}lrrr@{}}
\toprule
Dataset & SASRec & Fusion & Rel. gain \\
\midrule
Beauty & $0.0473\pm0.0008$ & $0.0817\pm0.0021$ & $+72.9\%$ \\
Toys & $0.0550\pm0.0022$ & $0.0999\pm0.0005$ & $+81.4\%$ \\
Sports & $0.0227\pm0.0004$ & $0.0516\pm0.0012$ & $+127.2\%$ \\
\bottomrule
\end{tabular*}
\caption{Original-SASRec control: three-seed mean $\pm$ s.d.; relative gains use unrounded means.}
\label{tab:weakbackbone}
\end{table}
\section{Reproducibility Assets}\label{supp:assets}
Code is available at \url{https://github.com/Double-wk/LIME-Rec}.
The accompanying code package contains a \texttt{code/} directory:
\texttt{lime\_rec/}, \texttt{scripts/}, \texttt{configs/}, \texttt{workflows/},
\texttt{prompts/}, compatibility patches, tests, and dependency files.
The \texttt{code/readme.md} gives startup commands, compact results, Figure~1,
and citation information. Raw datasets, checkpoints, embeddings, result/prediction
files, and Git history are excluded; the commands regenerate artifacts locally.
Data preparation uses \path{scripts/data/download_gram_data.py}.
Runtime requirements are in \texttt{pyproject.toml} (Python $\ge3.10$, PyTorch $2.5.1$);
GRAM uses the separate environment in Section~\ref{supp:reproduction}, and LM serving
and adaptation additionally require compatible vLLM, Transformers, and PEFT.
Original prompts and evaluation/paired-bootstrap code are included.
No training or inference experiment is rerun by preparing this package.

\subsection{Target Reproduction Settings}\label{supp:reproduction}
Table~\ref{tab:reproduction} records the implementation details for the matched
comparisons in Section~\ref{supp:targets}.
\begin{table}[H]
\centering\small
\setlength{\tabcolsep}{3pt}
\caption{Target reproduction settings.}\label{tab:reproduction}
\begin{tabular}{@{}p{0.26\columnwidth}p{0.68\columnwidth}@{}}
\toprule
Setting & Configuration \\
\midrule
GRAM revision & \path{7ac4d9272a57beed9df35c27ea34221f6e4a8fb1} \\
Training hardware & Two NVIDIA L20 GPUs; distributed training \\
Optimization & LR $10^{-3}$; batch $32$; gradient accumulation $2$; $30$ epochs; evaluation/checkpointing every $5$ epochs \\
Text & Prompt length $128$; all available item text \\
Identifier & Clusters/levels: Beauty $(128,7)$; Toys $(32,5)$; Sports $(32,7)$ \\
Neighbours & Beauty/Toys/Sports: $10$/$5$/$10$ \\
GRAM environment & Python 3.9.25; PyTorch 2.1.0+cu121; Transformers 4.26.0 \\
LIME-Rec environment & Python 3.11.13; PyTorch 2.5.1+cu124 (separate environment) \\
\bottomrule
\end{tabular}
\end{table}
The shared importer maps decoded semantic identifiers to raw IDs, parses trailing
gold/prediction/score TSV fields, and deduplicates distributed-sampler padding.
Alignment requires identical users, ordered histories, and targets; unmapped IDs
are rejected. Both exports use the same evaluator.

\subsection{Prompt Templates}\label{supp:prompts}
The following system prompts are the exact strings in the code package. Shared-pool
user messages attach candidates, history-membership flags, advertised tools, and
observed results. Limited-tool user messages serialize history, called/available
tools, remaining budget, and observed results as JSON.

\par\noindent\begin{minipage}{\columnwidth}
\textbf{Shared-pool tool-use controller.}\par
\begin{lstlisting}
You are a single recommendation controller. You receive opaque item IDs, history
membership flags, and only the outputs of tools you explicitly call. You never
receive a future item, ground-truth label, hit/miss signal, or raw item text.

Return exactly one JSON object. Either call one available, not-yet-called tool:
{"action":"call_tool","tool":"sasrec"}
or finish with at least ten unique candidate IDs:
{"action":"finish","ranking":["item1","item2","item3","item4","item5","item6","item7","item8","item9","item10"]}
Do not use tools outside the advertised list. Do not repeat a tool.

\end{lstlisting}
\end{minipage}\par\medskip

\par\noindent\begin{minipage}{\columnwidth}
\textbf{Shared-pool one-shot controller.}\par
\begin{lstlisting}
Rank the opaque candidate item IDs using only the supplied outputs of the three
expert tools and history-membership flags. No future interaction, evaluation
label, raw item text, or hidden tool score is available. Return exactly one JSON
object of the form {"action":"finish","ranking":[...]} containing at least ten
unique IDs from the candidate pool.

\end{lstlisting}
\end{minipage}\par\medskip

\par\noindent\begin{minipage}{\columnwidth}
\textbf{Limited-tool selector.}\par
\begin{lstlisting}
Select the next recommendation expert to query. sasrec models sequential behavior; itemcf models collaborative co-occurrence; semantic models frozen item-text similarity. You see only history and already queried expert results. Select one available tool. The environment will rank revealed items with fixed weighted score fusion after the call budget is exhausted. Return only JSON: {"tool":"sasrec"}, substituting an available tool name. Do not return a ranking or call an unavailable tool.
\end{lstlisting}
\end{minipage}\par\medskip

\subsection{LM Serving Configuration}\label{supp:serving}
Shared-pool serving uses vLLM 0.17.0's OpenAI-compatible interface. The recorded
Qwen3-8B deployment uses bf16 and tensor parallelism $1$ on one NVIDIA L20 (46~GB).
Fixed settings are temperature $0$, context length $32{,}768$, output budget $512$
tokens, at most $3$ tool calls and $2$ retries, and sampling seed $2027$.
Qwen thinking is disabled via \texttt{enable\_thinking=false}.
Validation freezes prompt hashes, model name, and checkpoint identity;
response caches are keyed by request hashes.

\end{document}